\documentclass[12pt,letterpaper]{article}
\usepackage{epsfig,rotating,setspace,latexsym,amsmath,epsf,amssymb,bm,theorem}
\usepackage{cite,appendix}

\usepackage{amsfonts}

\title{Degraded Compound
Multi-receiver Wiretap Channels\thanks{This work was supported by
NSF Grants CCF 04-47613, CCF 05-14846, CNS 07-16311 and CCF
07-29127, and presented in part at the 47th Annual Allerton
Conference on Communications, Control and Computing, Monticello,
IL, September 2009.}}

\author{Ersen Ekrem \qquad Sennur Ulukus \\
\normalsize Department of Electrical and Computer Engineering\\
\normalsize University of Maryland, College Park, MD 20742 \\
\normalsize {\it ersen@umd.edu} \qquad {\it ulukus@umd.edu}}

\newcommand{\bblambda}{\bm \Lambda}
\newcommand{\brho}{\bm \rho}
\newcommand{\bbdelta}{\bm \Delta}
\newcommand{\bbsigma}{\bm \Sigma}

\newcommand{\bbi}{{\mathbf{I}}}
\newcommand{\bzero}{{\mathbf{0}}}
\newcommand{\bbv}{{\mathbf{V}}}

\newcommand{\bw}{{\mathbf{w}}}
\newcommand{\bbh}{{\mathbf{H}}}

\newcommand{\bbm}{{\mathbf{M}}}

\newcommand{\bbk}{{\mathbf{K}}}
\newcommand{\bbz}{{\mathbf{Z}}}
\newcommand{\bbn}{{\mathbf{N}}}

\newcommand{\bba}{{\mathbf{A}}}
\newcommand{\bbd}{{\mathbf{D}}}

\newcommand{\bbt}{{\mathbf{T}}}
\newcommand{\bbb}{{\mathbf{B}}}

\newcommand{\bbs}{{\mathbf{S}}}
\newcommand{\bu}{{\mathbf{u}}}
\newcommand{\bbj}{{\mathbf{J}}}
\newcommand{\bbu}{{\mathbf{U}}}
\newcommand{\bx}{{\mathbf{x}}}
\newcommand{\bbx}{{\mathbf{X}}}

\newcommand{\bby}{{\mathbf{Y}}}

\newtheorem{Theo}{Theorem}

\newtheorem{Lem}{Lemma}

\newtheorem{Def}{Definition}

\begin{document}


\maketitle

\setstretch{1.2}

\begin{abstract}
In this paper, we study the degraded compound multi-receiver
wiretap channel. The degraded compound multi-receiver wiretap
channel consists of two groups of users and a group of
eavesdroppers, where, if we pick an arbitrary user from each group
of users and an arbitrary eavesdropper, they satisfy a certain
Markov chain. We study two different communication scenarios for
this channel. In the first scenario, the transmitter wants to send
a confidential message to users in the first (stronger) group and
a different confidential message to users in the second (weaker)
group, where both messages need to be kept confidential from the
eavesdroppers. For this scenario, we assume that there is only one
eavesdropper. We obtain the secrecy capacity region for the
general discrete memoryless channel model, the parallel channel
model, and the Gaussian parallel channel model. For the Gaussian
multiple-input multiple-output (MIMO) channel model, we obtain the
secrecy capacity region when there is only one user in the second
group. In the second scenario we study, the transmitter sends a
confidential message to users in the first group which needs to be
kept confidential from the second group of users and the
eavesdroppers. Furthermore, the transmitter sends a different
confidential message to users in the second group which needs to
be kept confidential only from the eavesdroppers. For this
scenario, we do not put any restriction on the number of
eavesdroppers. As in the first scenario, we obtain the secrecy
capacity region for the general discrete memoryless channel model,
the parallel channel model, and the Gaussian parallel channel
model. For the Gaussian MIMO channel model, we establish the
secrecy capacity region when there is only one user in the second
group.
\end{abstract}

\setstretch{1.2}
\newpage
\section{Introduction}

Information theoretic secrecy was initiated by Wyner in his
seminal work~\cite{Wyner}, where he considered the degraded
wiretap channel and established the capacity-equivocation rate
region of this degraded channel model. Later, Csiszar and Korner
generalized his result to arbitrary, not necessarily degraded,
wiretap channels in~\cite{Korner}. In recent years, multi-user
versions of the wiretap channel have attracted a considerable
amount of research interest; see for example references [3-21]
in~\cite{MIMO_BC_Secrecy}. Among all these extensions, two natural
extensions of the wiretap channel to the multi-user setting are
particularly of interest here: {\it secure broadcasting} and {\it
compound wiretap channels}.

{\it Secure broadcasting} refers to the situation where a
transmitter wants to communicate with several legitimate receivers
confidentially in the presence of an external eavesdropper. We
call this channel model the {\it multi-receiver wiretap channel}.
Since the underlying channel model without an eavesdropper is the
broadcast channel, which is not understood to the full extent even
for the two-user case, most works on {\it secure broadcasting}
have focused on some special classes of multi-receiver wiretap
channels, where these classes are identified by certain
degradation
orders~\cite{Broadcasting_Wornell,Khandani,Ekrem_Ulukus_Asilomar08,Ekrem_Ulukus_BC_Secrecy,El_Gamal_Secrecy}.
In particular,
\cite{Khandani,Ekrem_Ulukus_Asilomar08,Ekrem_Ulukus_BC_Secrecy}
consider the {\it degraded} multi-receiver wiretap channel, where
observations of all users and the eavesdropper satisfy a certain
Markov chain. In~\cite{Khandani}, the secrecy capacity region is
derived for the two-user case, and
in~\cite{Ekrem_Ulukus_Asilomar08,Ekrem_Ulukus_BC_Secrecy}, the
secrecy capacity region is established for an arbitrary number of
legitimate users. The importance of this result lies in the facts
that the Gaussian multi-receiver wiretap channel belongs to this
class, and the secrecy capacity region of the degraded
multi-receiver wiretap channel serves as a crucial step in
establishing the secrecy capacity region of the Gaussian
multiple-input multiple-output (MIMO) multi-receiver wiretap
channel~\cite{MIMO_BC_Secrecy}, though the latter channel is not
necessarily degraded. In~\cite{MIMO_BC_Secrecy}, besides proving
the secrecy capacity region of the Gaussian MIMO multi-receiver
wiretap channel, we also present new optimization results
regarding extremal properties of Gaussian random vectors, which we
generalize here.

Another extension of the wiretap channel that we are particularly
interested in here, is the {\it compound wiretap channel}. In
compound wiretap channels, there are a finite number of channel
states determining the channel transition probability. The channel
takes a certain fixed state for the entire duration of the
transmission, and the transmitter does not have any knowledge
about the channel state realization. Thus, the aim of the
transmitter is to ensure the secrecy of messages irrespective of
the channel state realization. In addition to this definition, the
compound wiretap channel admits another interpretation. Consider
the multi-receiver wiretap channel with several legitimate users
and many eavesdroppers, where the transmitter wants to transmit a
common confidential message to legitimate users while keeping all
of the eavesdroppers totally ignorant of the message. Since each
eavesdropper and legitimate user pair can be regarded as a
different channel state realization, this channel is equivalent to
a compound wiretap channel. Therefore, one can interpret a
compound wiretap channel as {\it multicasting} a common
confidential message to several legitimate receivers in the
presence of one or more eavesdroppers~\cite{Yingbin_Compound}. In
this work, we mostly refer to this interpretation, which is also
the reason why we classify the compound wiretap channel as an
extension of the wiretap channel to a multi-user setting.

Keeping this interpretation in mind, first works about the
compound wiretap channel are due to
Yamamoto~\cite{Yamamoto_Compound_1,Yamamoto_Compound_2}.
References~\cite{Yamamoto_Compound_1,Yamamoto_Compound_2} consider
the parallel wiretap channel with two sub-channels where each
sub-channel is wiretapped by a different eavesdropper.
References~\cite{Yamamoto_Compound_1,Yamamoto_Compound_2}
establish capacity-equivocation rate regions for the situation
where in each sub-channel, the legitimate receiver is less noisy
with respect to the eavesdropper of this sub-channel. Other works
which implicitly study the compound wiretap channel
are~\cite{Fading_Compound,Broadcasting_Wornell,Ekrem_Ulukus_Asilomar08,Ekrem_Ulukus_BC_Secrecy,El_Gamal_Secrecy},
where~\cite{Broadcasting_Wornell,Ekrem_Ulukus_Asilomar08,Ekrem_Ulukus_BC_Secrecy}
consider the transmission of a common confidential message to many
legitimate receivers in the presence of a single
eavesdropper,~\cite{El_Gamal_Secrecy} focuses on two legitimate
receivers one eavesdropper and one legitimate receiver two
eavesdroppers scenarios, and \cite{Fading_Compound} studies the
fading wiretap channel with many receivers.
Reference~\cite{Yingbin_Compound} considers the general discrete
compound wiretap channel and provides inner and outer bounds for
the secrecy capacity. In addition to these inner and outer
bounds,~\cite{Yingbin_Compound} also establishes the secrecy
capacity of the degraded compound wiretap channel as well as its
degraded Gaussian MIMO instance. Another work on the compound
wiretap channel is ~\cite{Tie_Liu_Compound_WT} where the secrecy
capacity of a class of non-degraded Gaussian parallel compound
wiretap channels is established.

\begin{figure}[t]
\begin{center}
\includegraphics[width=15cm]{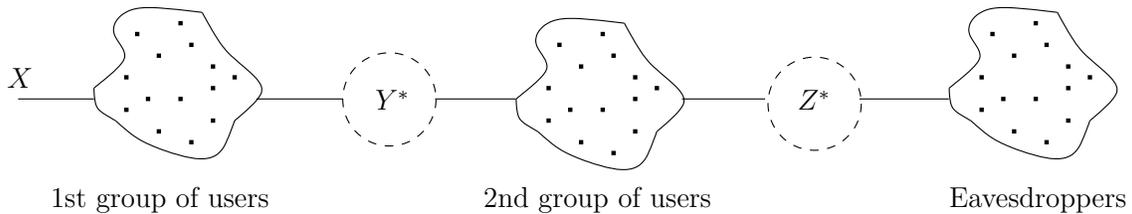}
\caption{The degraded compound multi-receiver wiretap channel.}
\label{fig_compound}
\end{center}
\end{figure}

In this work, we consider compound broadcast channels from a
secrecy point of view, which enables us to study the {\it secure
broadcasting} problem over {\it compound channels}. We note that
the current literature regarding the compound wiretap channel
considers the transmission of only one confidential message,
whereas here, we study the transmission of multiple confidential
messages, where each of these messages needs to be delivered to a
different group of users in perfect secrecy. Hereafter, we call
this channel model the {\it compound multi-receiver wiretap
channel} to emphasize the presence of more than one confidential
message. The compound multi-receiver wiretap channel we study here
consists of two groups of users and a group of eavesdroppers, as
shown in Figure~\ref{fig_compound}. We focus on a special class of
compound multi-receiver wiretap channels which exhibits a certain
degradation order. If we consider an arbitrary user from each
group and an arbitrary eavesdropper, they satisfy a certain Markov
chain. In particular, we assume that there exist two fictitious
users. The first fictitious user is degraded with respect to any
user from the first group, and any user from the second group is
degraded with respect to the first fictitious user. There exists a
similar degradedness structure for the second fictitious user in
the sense that it is degraded with respect to any user from the
second group, and any eavesdropper is degraded with respect to it.
Without eavesdroppers, this channel model reduces to the degraded
compound broadcast channel studied in~\cite{Tie_Liu_Compound}.
Adapting their terminology, we call our channel model the {\it
degraded compound multi-receiver wiretap channel}. Here, we
consider the general discrete memoryless version of the degraded
compound multi-receiver wiretap channel as well as its
specializations to the parallel degraded compound multi-receiver
wiretap channel, the Gaussian parallel degraded compound
multi-receiver wiretap channel, and the Gaussian MIMO degraded
compound multi-receiver wiretap channel. We study two different
communication scenarios for each version of the degraded compound
multi-receiver wiretap channel model.

\begin{figure}[t]
\begin{center}
\includegraphics[width=12.5cm]{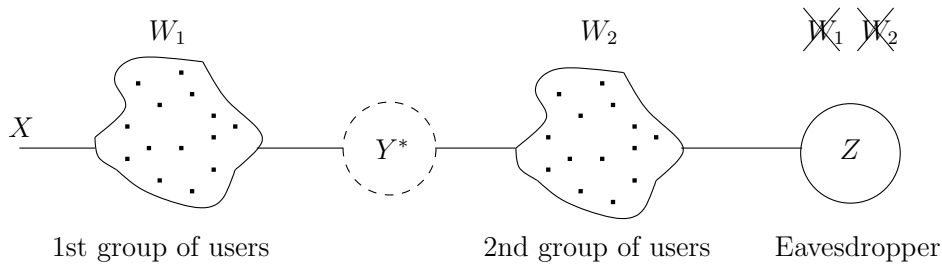}
\caption{The first scenario for the degraded compound
multi-receiver wiretap channel.} \label{fig_compound_scenario_1}
\end{center}
\end{figure}

In the first scenario, which is illustrated in
Figure~\ref{fig_compound_scenario_1}, the transmitter wants to
send a confidential message to users in the first group, and a
different confidential message to users in the second group, where
both messages need to be kept confidential from the eavesdroppers.
For this scenario, we assume that there exists only one
eavesdropper and obtain the secrecy capacity region in a
single-letter form. While obtaining this result, the presence of
the fictitious user between the two groups of users plays a
crucial role in the converse proof by providing a conditional
independence structure in the channel, which enables us to define
an auxiliary random variable that yields a tight outer bound.
After establishing single-letter expressions for the secrecy
capacity region, we consider the parallel degraded compound
multi-receiver wiretap channel. For the parallel degraded compound
multi-receiver wiretap channel, we obtain the secrecy capacity
region in a single-letter form as well. Though the general
discrete memoryless degraded compound multi-receiver wiretap
channel encompasses the parallel degraded compound multi-receiver
wiretap channel as a special case, we still need a converse proof
to establish the optimality of independent signalling in each
sub-channel. After we obtain the secrecy capacity region of the
parallel degraded compound multi-receiver wiretap channel, we
consider the Gaussian parallel degraded compound multi-receiver
wiretap channel. In particular, we evaluate the secrecy capacity
region of the parallel degraded compound multi-receiver wiretap
channel for the Gaussian case, which is tantamount to finding the
optimal joint distribution of auxiliary random variables and
channel inputs, which is shown to be Gaussian. We accomplish this
by using Costa's entropy power inequality~\cite{Costa_EPI}.
Finally, we consider the Gaussian MIMO degraded compound
multi-receiver wiretap channel, and evaluate its secrecy capacity
region when there is only one user in the second group. We show
the optimality of a jointly Gaussian distribution for auxiliary
random variables and channel inputs by generalizing our
optimization results in~\cite{MIMO_BC_Secrecy}.

\begin{figure}[htp]
\begin{center}
\includegraphics[width=15cm]{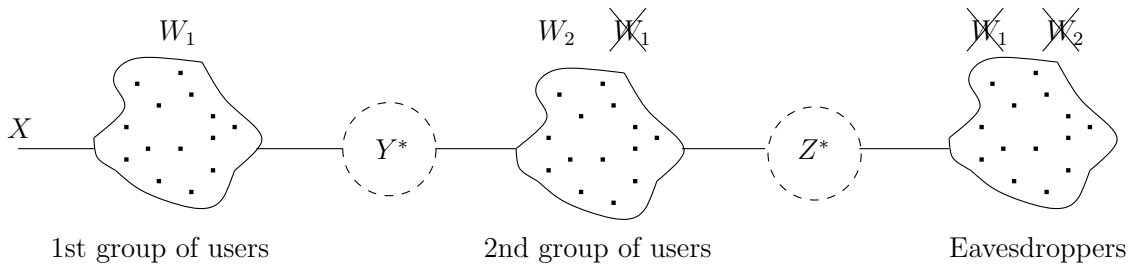}
\caption{The second scenario for the degraded compound
multi-receiver wiretap channel.} \label{fig_compound_scenario_2}
\end{center}
\end{figure}

In the second scenario we study here, which is illustrated in
Figure~\ref{fig_compound_scenario_2}, the transmitter wants to
send a confidential message to users in the first group which
needs to be kept confidential from users in the second group and
eavesdroppers. Moreover, the transmitter sends a different
confidential message to users in the second group, which needs to
be kept confidential from the eavesdroppers. If there were only
one user in each group and one eavesdropper, this channel model
would reduce to the channel model that was studied
in~\cite{Vector_Costa_EPI}. However, here, there are an arbitrary
number of users in each group and an arbitrary number of
eavesdroppers. Hence, our model can be viewed as a generalization
of~\cite{Vector_Costa_EPI} to a compound setting. Adapting their
terminology, we call this channel model the {\it degraded compound
multi-receiver wiretap channel with layered messages}. We first
obtain the secrecy capacity region in a single-letter form for a
general discrete memoryless setting, where again the presence of
fictitious users plays a key role in the converse proof. Next, we
consider the parallel degraded compound multi-receiver wiretap
channel with layered messages and establish its secrecy capacity
region in a single-letter form. In this case as well, we provide
the converse proof which is again necessary to show the optimality
of independent signalling in each sub-channel. After we obtain the
secrecy capacity region of the parallel degraded compound
multi-receiver wiretap channel with layered messages, we evaluate
it for the Gaussian parallel degraded compound multi-receiver
wiretap channel with layered messages by showing the optimality of
a jointly Gaussian distribution for auxiliary random variables and
channel inputs. For that purpose, we again use Costa's entropy
power inequality~\cite{Costa_EPI}. Finally, we consider the
Gaussian MIMO degraded compound multi-receiver wiretap channel
with layered messages, and evaluate its secrecy capacity region
when there is only one user in the second group. To this end, we
show that jointly Gaussian auxiliary random variables and channel
inputs are optimal by extending our optimization results
in~\cite{MIMO_BC_Secrecy}.

\section{System Model}

In this paper, we consider the degraded compound multi-receiver
wiretap channel, see Figure~\ref{fig_compound}, which consists of
two groups of users and a group of eavesdroppers. There are $K_1$
users in the first group, $K_2$ users in the second group, and
$K_Z$ eavesdroppers. The channel is assumed to be memoryless with
a transition probability
\begin{align}
p(y_1^1,\ldots,y_{K_1}^1,y_{1}^2,\ldots,y_{K_2}^{2},z_1,\ldots,z_{K_Z}|x)
\end{align}
where $X\in \mathcal{X}$ is the channel input,
$Y_j^1\in\mathcal{Y}_j^1$ is the channel output of the $j$th user
in the first group, $j=1,\ldots,K_1$, $Y_k^2\in\mathcal{Y}_k^2$ is
the channel output of the $k$th user in the second group,
$k=1,\ldots,K_2$, and $Z_t\in\mathcal{Z}_t$ is the channel output
of the $t$th eavesdropper, $t=1,\ldots,K_Z$.

We assume that there exist two fictitious users with observations
$Y^*\in\mathcal{Y}^*, Z^*\in\mathcal{Z}^*$ such that they satisfy
the Markov chain
\begin{align}
X\rightarrow Y_{j}^1 \rightarrow Y^* \rightarrow Y_k^2 \rightarrow
Z^* \rightarrow Z_t,\quad \forall (j,k,t) \label{our_Markov_chain}
\end{align}
This Markov chain is the reason why we call this channel model the
{\it degraded compound multi-receiver wiretap channel}. Actually,
there is a slight inexactness in the terminology here because the
Markov chain in (\ref{our_Markov_chain}) is more restrictive than
the Markov chain
\begin{align}
X\rightarrow Y_{j}^1 \rightarrow Y_k^2  \rightarrow Z_t,\quad
\forall (j,k,t) \label{broader_Markov_chain}
\end{align}
and it might be more natural to define the degradedness of the
compound multi-receiver wiretap channel by the Markov chain in
(\ref{broader_Markov_chain}). However, in this work, we adapt the
terminology of the previous work on compound broadcast
channels~\cite{Tie_Liu_Compound}, and call the channel satisfying
(\ref{our_Markov_chain}) the degraded compound multi-receiver
wiretap channel. Finally, we note that when there are no
eavesdroppers, this channel reduces to the degraded compound
broadcast channel that was studied in~\cite{Tie_Liu_Compound}.

\subsection{Parallel Degraded Compound Multi-receiver Wiretap Channels}

The parallel degraded compound multi-receiver wiretap channel,
where each user's and each eavesdropper's channel consists of $L$
independent sub-channels, i.e.,
\begin{align}
Y_j^1&=(Y_{j1}^1,\ldots,Y_{jL}^1),\quad j=1,\ldots,K_1\\
Y_k^2&=(Y_{k1}^2,\ldots,Y_{kL}^2),\quad k=1,\ldots,K_2 \\
Z_t&=(Z_{t1},\ldots,Z_{tL}),\quad t=1,\ldots,K_Z
\end{align}
has the following overall transition probability
\begin{align}
p(y_1^1,\ldots,y_{K_1}^1,y_{1}^2,\ldots,y_{K_2}^{2},z_1,\ldots,z_{K_Z}|x)=\prod_{\ell=1}^L
p(y_{1\ell}^1,\ldots,y_{K_1\ell}^1,y_{1\ell}^2,\ldots,y_{K_2\ell}^{2},z_{1\ell},\ldots,z_{K_Z\ell}|x_{\ell})
\end{align}
where $X_{\ell},~\ell=1,\ldots,L,$ is the $\ell$th sub-channel's
input. We define the degradedness of the parallel compound
multi-receiver wiretap channel in a similar fashion. In
particular, we call a parallel compound multi-receiver wiretap
channel degraded, if there exist two sequences of random variables
\begin{align}
Y^*&=(Y_{1}^*,\ldots,Y^*_L) \\
Z^*&=(Z_{1}^*,\ldots,Z^*_L)
\end{align}
which satisfy Markov chains
\begin{align}
X_{\ell}\rightarrow Y_{j\ell}^{1} \rightarrow Y_{\ell}^*
\rightarrow Y_{k\ell}^2 \rightarrow Z_{\ell}^* \rightarrow
Z_{t\ell},\quad \forall(j,k,t,\ell) \label{our_Markov_chain_1}
\end{align}

\subsection{Gaussian Parallel Degraded Compound Multi-receiver Wiretap Channels}

The Gaussian parallel compound multi-receiver wiretap channel is
defined by
\begin{align}
\bby^1_{j}&=\bbx+\bbn^1_j,\quad j=1,\ldots,K_1 \\
\bby^2_k&=\bbx+\bbn_k^2,\quad k=1,\ldots,K_2 \\
\bbz_t&=\bbx+\bbn_t^Z,\quad t=1,\ldots,K_Z
\end{align}
where all column vectors
$\{\bby_j^1\}_{j=1}^{K_1},\{\bby_k^1\}_{k=1}^{K_2},\{\bbz_t\}_{t=1}^{K_Z},\bbx,
\{\bbn^1_j\}_{j=1}^{K_1},\{\bbn^2_k\}_{k=1}^{K_2},\{\bbn_t^Z\}_{t=1}^{K_Z}$
are of dimensions $L\times 1$.
$\{\bbn^1_j\}_{j=1}^{K_1},\{\bbn^2_k\}_{k=1}^{K_2},\{\bbn_t^Z\}_{t=1}^{K_Z}$
are Gaussian random vectors with diagonal covariance matrices
$\{\bblambda_j^1\}_{j=1}^{K_1},\{\bblambda_k^2\}_{j=1}^{K_2},\{\bblambda_t^Z\}_{t=1}^{K_Z}$,
respectively. The channel input $\bbx$ is subject to a trace
constraint as
\begin{align}
E\left[\bbx^\top \bbx\right]={\rm tr} \left(E\left[\bbx
\bbx^\top\right]\right) \leq P
\end{align}

In this paper, we will be interested in Gaussian parallel {\it
degraded} compound multi-receiver wiretap channels which means
that the covariance matrices satisfy the following order
\begin{align}
\bblambda_j^1\preceq \bblambda_k^2 \preceq \bblambda_t^Z,\quad
\forall(j,k,t) \label{positive_defi_order}
\end{align}
Since noise covariance matrices are diagonal, the order in
(\ref{positive_defi_order}) implies
\begin{align}
\Lambda_{j,\ell\ell}^1\leq  \Lambda_{k,\ell\ell}^2 \leq
\Lambda_{t,\ell\ell}^Z,\quad \forall (j,k,t,\ell)
\label{positive_defi_order_elementwise}
\end{align}
where $\Lambda_{j,\ell\ell}^1, \Lambda_{k,\ell\ell}^2
,\Lambda_{t,\ell\ell}^Z$ denote the $\ell$th diagonal element of
$\bblambda_j^1,\bblambda_k^2,\bblambda_t^Z$, respectively.

The diagonality of noise covariance matrices also ensures the
existence of diagonal matrices $\bblambda_Y^*$ and $\bblambda^*_Z$
such that
\begin{align}
\bblambda_j^1 \preceq \bblambda_Y^* \preceq \bblambda_k^2 \preceq
\bblambda_Z^* \preceq \bblambda_t^Z,\quad \forall (k,j,t)
\label{parallel_is_ours}
\end{align}
For example, we can select $\bblambda_Y^*$ as
$\Lambda_{Y,\ell\ell}^*=\max_{j=1,\ldots,K_1}\Lambda_{j,\ell\ell}^{1}$
which already satisfies (\ref{parallel_is_ours}) because of
$\max_{j=1,\ldots,K_1}\Lambda_{j,\ell\ell}^{1}\leq
\min_{k=1,\ldots,K_2}\Lambda_{k,\ell\ell}^2$ which is due to
(\ref{positive_defi_order_elementwise}). Similarly, we can select
$\bblambda_Z^*$. Thus, for Gaussian parallel compound
multi-receiver channels, the two possible ways of defining
degradedness, i.e., (\ref{our_Markov_chain}) and
(\ref{broader_Markov_chain}), are equivalent due to the
equivalence of (\ref{positive_defi_order}) and
(\ref{parallel_is_ours}).

\subsection{Gaussian MIMO Degraded Compound Multi-receiver Wiretap Channels}

The Gaussian MIMO degraded compound multi-receiver wiretap channel
is defined by
\begin{align}
\bby_j^1&=\bbx+\bbn_j^1,\quad j=1,\ldots,K_1 \label{MIMO_channel_def_1} \\
\bby_k^2&=\bbx+\bbn_k^2,\quad k=1,\ldots,K_2\label{MIMO_channel_def_2} \\
\bbz_t&=\bbx+\bbn_t^Z,\quad t=1,\ldots,K_Z
\label{MIMO_channel_def_3}
\end{align}
where all column vectors
$\{\bby_j^1\}_{j=1}^{K_1},\{\bby_k^2\}_{k=1}^{K_2},\{\bbz_t\}_{t=1}^{K_Z},\bbx,\{\bbn_j^1\}_{j=1}^{K_1},
\{\bbn_k^2\}_{k=1}^{K_2},\{\bbn_t^Z\}_{t=1}^{K_Z}$ are of
dimensions $M\times 1$. $\{\bbn_j^1\}_{j=1}^{K_1},
\{\bbn_k^2\}_{k=1}^{K_2},\{\bbn_t^Z\}_{t=1}^{K_Z}$ are Gaussian
random vectors with covariance matrices
$\{\bbsigma_j^1\}_{j=1}^{K_1},\{\bbsigma_k^2\}_{k=1}^{K_2},\{\bbsigma_t^Z\}_{t=1}^{K_Z}$,
respectively. Unlike in the case of Gaussian parallel channels,
these covariance matrices are not necessarily diagonal. The
channel input $\bbx$ is subject to a covariance constraint
\begin{align}
E\left[\bbx\bbx^\top\right] \preceq \bbs
\label{covariance_constraint}
\end{align}
where $\bbs \succ \bzero$.

In this paper, we study Gaussian MIMO {\it degraded} compound
multi-receiver wiretap channels for which there exist covariance
matrices $\bbsigma_Y^*$ and $\bbsigma_Z^*$ such that
\begin{align}
\bbsigma_j^1 \preceq\bbsigma_Y^* \preceq \bbsigma_k^2 \preceq
\bbsigma_Z^* \preceq \bbsigma_t^Z,\quad \forall (j,k,t)
\label{order_of_us}
\end{align}
We note that the order in (\ref{order_of_us}), by which we define
the degradedness, is more restrictive than the other possible
order that can be used to define the degradedness, i.e.,
\begin{align}
\bbsigma_j^1  \preceq \bbsigma_k^2  \preceq \bbsigma_t^Z,\quad
\forall (j,k,t) \label{order_of_broader}
\end{align}
In~\cite{Tie_Liu_Compound}, a specific numerical example is
provided to show that the order in (\ref{order_of_broader})
strictly subsumes the one in (\ref{order_of_us}).

\subsection{Comments on Gaussian MIMO Degraded Compound Multi-receiver Wiretap Channels}
 We provide some comments about the way we define the Gaussian MIMO degraded compound
 multi-receiver wiretap channel. The first one is about the
 covariance constraint in (\ref{covariance_constraint}). Though it is more common to define capacity regions
 under a total power constraint, i.e., ${\rm tr}\left(E\left[\bbx \bbx^{\top}\right]\right) \leq
 P$, the covariance constraint in
 (\ref{covariance_constraint}) is more general and it
 subsumes the total power constraint as a special case~\cite{Shamai_MIMO}.
In particular, if we denote the secrecy capacity region under the
constraint in (\ref{covariance_constraint}) by $C(\bbs)$, then the
secrecy capacity region under the trace constraint, ${\rm
tr}\left(E\left[\bbx\bbx^{\top}\right]\right)\leq P$, can be
written as~\cite{Shamai_MIMO}
\begin{align}
C^{\rm trace}(P)=\bigcup_{\bbs:{\rm tr}(\bbs)\leq P} C(\bbs)
\end{align}

The second comment is about our assumption that $\bbs$ is strictly
positive definite. This assumption does not lead to any loss of
generality because for any Gaussian MIMO compound multi-receiver
wiretap channel with a positive semi-definite covariance
constraint, i.e., $\bbs \succeq \bzero$ and $|\bbs|=0$, we can
always construct an equivalent channel with the constraint
$E\left[\bbx \bbx^{\top}\right]\preceq \bbs^{\prime}$ where
$\bbs^{\prime}\succ \bzero$ (see Lemma~2 of \cite{Shamai_MIMO}),
which has the same secrecy capacity region.

The last comment is about the assumption that the transmitter and
all receivers have the same number of antennas. This assumption is
implicit in the channel definition, see
(\ref{MIMO_channel_def_1})-(\ref{MIMO_channel_def_3}), and also in
the definition of degradedness, see (\ref{order_of_us}). However,
we can extend the definition of the Gaussian MIMO degraded
compound multi-receiver wiretap channel to include the cases where
the number of transmit antennas and the number of receive antennas
at each receiver are not necessarily the same. To this end, we
first introduce the following channel model
\begin{align}
\bby_j^1&=\bbh_j^1 \bbx+\bbn_j^1,\quad j=1,\ldots,K_1\label{MIMO_channel_def_again_1}\\
\bby_k^2&=\bbh_k^2 \bbx+\bbn_k^2,\quad k=1,\ldots,K_2\label{MIMO_channel_def_again_2}\\
\bbz_t&=\bbh_t^Z \bbx+\bbn_t^Z,\quad
t=1,\ldots,K_Z\label{MIMO_channel_def_again_3}
\end{align}
where $\bbh_j^1,\bbh_k^2,\bbh_t^Z $ are the channel matrices of
sizes $r_j^1 \times t, r_k^2 \times t, r_t^Z \times t$,
respectively, and $\bbx$ is of size $t\times 1$. The channel
outputs $\bby_j^1,\bby_k^2,\bbz_t$ are of sizes $r_j^1 \times 1,
r_k^2 \times 1, r_t^Z \times 1$, respectively. The Gaussian noise
vectors $\bbn_j^1,\bbn_k^2,\bbn_t^Z$ are assumed to have identity
covariance matrices.

To define degradedness for the channel model given in
(\ref{MIMO_channel_def_again_1})-(\ref{MIMO_channel_def_again_3}),
we need the following definition from~\cite{Tie_Liu_Compound}: A
receive vector $\bby_a=\bbh_a\bbx+\bbn_a$ of size $r_a\times 1$ is
said to be degraded with respect to $\bby_b=\bbh_b\bbx+\bbn_b$ of
size $r_b\times 1$, if there exists a matrix $\bbd$ of size
$r_a\times r_b$ such that $\bbd\bbh_b=\bbh_a$ and $\bbd\bbd^\top
\preceq \bbi$. Using this equivalent definition of degradedness,
we now give the equivalent definition of degradedness for the
channel model in
(\ref{MIMO_channel_def_again_1})-(\ref{MIMO_channel_def_again_3}).
To this end, we first introduce two fictitious users with
observations $\bby^*$ and $\bbz^*,$ which are given by
\begin{align}
\bby^*&=\bbh^*_Y\bbx +\bbn_Y^* \\
\bbz^*&=\bbh^*_Z\bbx+\bbn_Z^*
\end{align}
The Gaussian MIMO compound multi-receiver wiretap channel in
(\ref{MIMO_channel_def_again_1})-(\ref{MIMO_channel_def_again_3})
is said to be degraded if the following two conditions hold: i)
$\bby^*$ is degraded with respect to any user from the first
group, and any user from the second group is degraded with respect
to $\bby^*$, and ii) $\bbz^*$ is degraded with respect to any user
from the second group, and any eavesdropper is degraded with
respect to $\bbz^*$, where degradedness here is with respect to
the definition given above.

In the rest of the paper, we consider the channel model given in
(\ref{MIMO_channel_def_1})-(\ref{MIMO_channel_def_3}) instead of
the channel model given in
(\ref{MIMO_channel_def_again_1})-(\ref{MIMO_channel_def_again_3}),
which is more general. However, if we establish the secrecy
capacity region for the Gaussian MIMO degraded compound
multi-receiver wiretap channel defined by
(\ref{MIMO_channel_def_1})-(\ref{MIMO_channel_def_3}), we can also
obtain the secrecy capacity region for the Gaussian MIMO degraded
compound multi-receiver wiretap channel defined by
(\ref{MIMO_channel_def_again_1})-(\ref{MIMO_channel_def_again_3})
using the analysis carried out in
Section~V~of~\cite{Tie_Liu_Compound} and Section 7.1 of
\cite{MIMO_BC_Secrecy}. Thus, focusing on the channel model in
(\ref{MIMO_channel_def_1})-(\ref{MIMO_channel_def_3}) does not
result in any loss of generality.

\section{Problem Statement and Main Results}

In this paper, we consider two different communication scenarios
for the degraded compound multi-receiver wiretap channel.

\subsection{The First Scenario: External Eavesdroppers}

In the first scenario, the transmitter wants to send a
confidential message to users in the first group and a different
confidential message to users in the second group, where both
messages need to be kept confidential from the eavesdroppers. In
this case, we assume that there is only one eavesdropper, i.e.,
$K_Z=1$. The graphical illustration of the first scenario is given
in Figure~\ref{fig_compound_scenario_1}.

An $(n,2^{nR_1},2^{nR_2})$ code for the first scenario consists of
two message sets $\mathcal{W}_1=\{1,\ldots,\break2^{nR_1}\},
\mathcal{W}_2=\{1,\ldots,2^{nR_2}\},$ an encoder
$f:\mathcal{W}_1\times \mathcal{W}_2\rightarrow \mathcal{X}^n$,
one decoder for each legitimate user in the first group
$g^1_{j}:\mathcal{Y}^{1,n}_j\rightarrow \mathcal{W}_1$,
$j=1,\ldots,K_1$, and one decoder for each legitimate user in the
second group $g^2_{k}:\mathcal{Y}^{2,n}_k\rightarrow
\mathcal{W}_2,~k=1,\ldots,K_2$. The probability of error is
defined as
\begin{align}
P_e^n=\max \left\{P_e^{1,n},P_{e}^{2,n}\right\}
\end{align}
where $P_e^{1,n}$ and $P_e^{2,n}$ are given by
\begin{align}
P_e^{1,n}&=\max_{
j\in\{1,\ldots,K_1\}}\Pr\left[g^1_j\big(Y^{1,n}_j\big)\neq
W_1\right]\\
P_e^{2,n}&=\max_{
k\in\{1,\ldots,K_2\}}\Pr\left[g^2_k\big(Y^{2,n}_k\big)\neq
W_2\right]
\end{align}
A secrecy rate pair $(R_1,R_2)$ is said to be achievable if there
exists an $(n,2^{nR_1},2^{nR_2})$ code which has
$\lim_{n\rightarrow \infty}P_e^n=0$ and
\begin{align}
\lim_{n\rightarrow \infty} \frac{1}{n}I(W_1,W_2;Z^n)=0
\label{perfect_secrecy}
\end{align}
where we dropped the subscript of $Z_t$ since $K_Z=1$. We note
that (\ref{perfect_secrecy}) implies
\begin{align}
\lim_{n\rightarrow \infty} \frac{1}{n}I(W_1;Z^n)=0\qquad
\textrm{and}\qquad \lim_{n\rightarrow \infty}
\frac{1}{n}I(W_2;Z^n)=0
\end{align}
From these definitions, it is clear that we are only interested in
perfect secrecy rates of the channel. The secrecy capacity region
is defined as the closure of all achievable secrecy rate pairs. A
single-letter characterization of the secrecy capacity region is
given as follows.

\begin{Theo}
\label{theorem_discrete} The secrecy capacity region of the
degraded compound multi-receiver wiretap channel is given by the
union of rate pairs $(R_1,R_2)$ satisfying
\begin{align}
R_1&\leq \min_{j=1,\ldots,K_1} I(X;Y^1_j|U,Z)\\
R_2 & \leq \min_{k=1,\ldots,K_2} I(U;Y^2_k|Z)
\end{align}
where the union is over all $(U,X)$ such that
\begin{align}
U\rightarrow X\rightarrow Y_j^1 \rightarrow Y^* \rightarrow
Y_k^2\rightarrow Z
\end{align}
for any $(j,k)$ pair.
\end{Theo}
Showing the achievability of this region is rather standard, thus
is omitted here. We provide the converse proof in
Appendix~\ref{proof_of_theorem_discrete}. The presence of the
fictitious user with observation $Y^*$ proves to be crucial in the
converse proof. Essentially, it brings a conditional independence
structure to the channel, which enables us to define the auxiliary
random variable $U$, which, in turn, provides the converse proof.

As a side note, if we disable the eavesdropper by setting
$Z=\phi$, the region in Theorem~\ref{theorem_discrete} reduces to
the capacity region of the underlying degraded compound broadcast
channel which was established in~\cite{Tie_Liu_Compound}.

\subsubsection{Parallel Degraded Compound Multi-Receiver Wiretap Channels}

In the upcoming section, we will consider the Gaussian parallel
degraded compound multi-receiver wiretap channel. For that
purpose, here, we provide the secrecy capacity region of the
parallel degraded compound multi-receiver wiretap channel in a
single-letter form.
\begin{Theo}
\label{theorem_discrete_parallel} The secrecy capacity region of
the parallel degraded compound multi-receiver wiretap channel is
given by the union of rate pairs $(R_1,R_2)$ satisfying
\begin{align}
R_1 &\leq \min_{j=1,\ldots,K_1} \sum_{\ell=1}^L
I(X_{\ell};Y_{j\ell}^1|U_{\ell},Z_{\ell}) \label{theorem_discrete_parallel_R1} \\
R_2 &\leq \min_{k=1,\ldots,K_2} \sum_{\ell=1}^L
I(U_{\ell};Y_{k\ell}^2|Z_{\ell})
\label{theorem_discrete_parallel_R2}
\end{align}
where the union is over all distributions of the form
$\prod_{\ell=1}^L p(u_{\ell},x_{\ell})$ such that
\begin{align}
U_{\ell}\rightarrow X_{\ell} \rightarrow Y_{j\ell}^1 \rightarrow
Y_{\ell}^* \rightarrow Y_{k\ell}^2 \rightarrow Z_{\ell}
\end{align}
for any $(j,k,\ell)$ triple.
\end{Theo}
Though Theorem~\ref{theorem_discrete} provides the secrecy
capacity region for a rather general channel model including the
parallel degraded compound multi-receiver channel as a special
case, we still need a converse proof to show that the region in
Theorem~\ref{theorem_discrete} reduces to the region in
Theorem~\ref{theorem_discrete_parallel} for parallel channels. In
other words, we still need to show the optimality of independent
signalling on each sub-channel. This proof is provided in
Appendix~\ref{proof_of_theorem_discrete_parallel}.

\subsubsection{Gaussian Parallel Degraded Compound Multi-Receiver Wiretap
Channels}

We now obtain the secrecy capacity region of the parallel Gaussian
degraded compound multi-receiver wiretap channel. To that end, we
need to evaluate the region given in
Theorem~\ref{theorem_discrete_parallel}, i.e., we need to find the
optimal joint distribution $\prod_{\ell=1}^L
p(u_{\ell},x_{\ell})$. We first introduce the following theorem
which will be instrumental in evaluating the region in
Theorem~\ref{theorem_discrete_parallel} for Gaussian parallel
channels.

\begin{Theo}
\label{theorem_parallel_Gauss_optimization} Let $N_1,N^*,N_2,N_Z$
be zero-mean Gaussian random variables with variances
$\sigma_1^2,\sigma_*^2,\sigma_2^2,\sigma_Z^2,$ respectively, where
\begin{align}
\sigma_1^2\leq \sigma_*^2\leq \sigma_2^2\leq \sigma_Z^2
\label{dummy_order}
\end{align}
Let $(U,X)$ be an arbitrarily dependent random variable pair,
which is independent of \break $(N_1,N^*,N_2,N_Z)$, and the
second-moment of $X$ be constrained as $E\left[X^2\right] \leq P$.
Then, for any feasible $(U,X)$, we can find a $P^*\leq P$ such
that
\begin{align}
h(X+N_Z|U)-h(X+N^*|U)&=\frac{1}{2} \log
\frac{P^*+\sigma_Z^2}{P^*+\sigma_*^2}
\end{align}
and
\begin{align}
h(X+N_Z|U)-h(X+N_1|U)&\geq \frac{1}{2} \log
\frac{P^*+\sigma_Z^2}{P^*+\sigma_1^2}\label{inequality_1}\\
h(X+N_Z|U)-h(X+N_2|U)&\leq \frac{1}{2} \log
\frac{P^*+\sigma_Z^2}{P^*+\sigma_2^2} \label{inequality_2}
\end{align}
for any $(\sigma_1^2,\sigma_2^2)$ satisfying the order in
(\ref{dummy_order}).
\end{Theo}
Costa's entropy power inequality~\cite{Costa_EPI} plays a key role
in the proof of this theorem. The proof of this theorem is
provided in
Appendix~\ref{proof_of_theorem_parallel_Gauss_optimization}.

We are now ready to establish the secrecy capacity region of the
Gaussian parallel degraded compound multi-receiver wiretap
channel.
\begin{Theo}
\label{theorem_parallel_Gaussian} The secrecy capacity region of
the Gaussian parallel degraded compound multi-receiver wiretap
channel is given by the union of rate pairs $(R_1,R_2)$ satisfying
\begin{align}
R_1 &\leq \min_{j=1,\ldots,K_1} \sum_{\ell=1}^L \frac{1}{2} \log
\left(1+\frac{\beta_{\ell}P_{\ell}}{\Lambda^1_{j,\ell\ell}}\right)
-\frac{1}{2} \log
\left(1+\frac{\beta_{\ell}P_{\ell}}{\Lambda_{Z,\ell\ell}}\right)
\label{parallel_Gaussian_R1}
\\
R_2 & \leq \min_{k=1,\ldots,K_2} \sum_{\ell=1}^L \frac{1}{2} \log
\left(1+\frac{\bar{\beta}_{\ell}P_{\ell}}{\beta_{\ell}P_{\ell}+\Lambda^2_{k,\ell\ell}}\right)
-\frac{1}{2} \log
\left(1+\frac{\bar{\beta}_{\ell}P_{\ell}}{\beta_{\ell}P_{\ell}+\Lambda_{Z,\ell\ell}}\right)
\label{parallel_Gaussian_R2}
\end{align}
where the union is over all $\{P_{\ell}\}_{\ell=1}^L$ such that $
\sum_{\ell=1}^L P_{\ell}=P$ and
$\bar{\beta}_{\ell}=1-\beta_{\ell}\in[0,1],~\ell=1,\ldots,L $.
\end{Theo}
The proof of this theorem is provided in
Appendix~\ref{proof_of_theorem_parallel_Gaussian}. Here,
$P_{\ell}$ denotes the part of the total available power $P$ which
is devoted to the transmission in the $\ell$th sub-channel.
Furthermore, $\beta_{\ell}$ denotes the fraction of the power
$P_{\ell}$ of the $\ell$th sub-channel spent for the transmission
to users in the first group.

\subsubsection{Gaussian MIMO Degraded Compound Multi-receiver Wiretap Channels}

In this section, we first obtain the secrecy capacity region of
the Gaussian MIMO degraded compound multi-receiver wiretap channel
when $K_2=1$, and then partially characterize the secrecy capacity
region for the case $K_2>1$.  To that end, we need to evaluate the
region given in Theorem~\ref{theorem_discrete}. In other words, we
need to find the optimal random variable pair $(U,\bbx)$. We are
able to do this for the entire capacity region when there is only
one user in the second group, i.e., $K_2=1$. For this, we need the
following theorem.

\begin{Theo}
\label{theorem_mimo_optimization} Let $(\bbn_1,\bbn^*,\bbn_Z)$ be
zero-mean Gaussian random vectors with covariance matrices
$\bbsigma_1,\bbsigma^*,\bbsigma_Z$, respectively, where
\begin{align}
\bbsigma_1 \preceq \bbsigma^* \preceq \bbsigma_Z
\label{dummy_order_1}
\end{align}
Let $(U,\bbx)$ be arbitrarily dependent random vector, which is
independent of $(\bbn_1,\bbn^*,\bbn_Z)$, and let the second moment
of $\bbx$ be constrained as $E\left[\bbx \bbx^{\top}\right]\preceq
\bbs$. Then, for any feasible $(U,\bbx)$, we can find a positive
semi-definite matrix $\bbk^*$ such that $\bbk^*\preceq \bbs$, and
it satisfies
\begin{align}
h(\bbx+\bbn_Z|U)-h(\bbx+\bbn^*|U)=\frac{1}{2} \log
\frac{|\bbk^*+\bbsigma_Z|}{|\bbk^*+\bbsigma^*|}
\label{mimo_equality}
\end{align}
and
\begin{align}
h(\bbx+\bbn_Z|U)-h(\bbx+\bbn_1|U)\geq \frac{1}{2} \log
\frac{|\bbk^*+\bbsigma_Z|}{|\bbk^*+\bbsigma_1|}
\label{mimo_inequality}
\end{align}
for any $\bbsigma_1$ satisfying the order in
(\ref{dummy_order_1}).
\end{Theo}
The proof of this theorem can be found in~\cite{MIMO_BC_Secrecy}.
Using this theorem, we can establish the secrecy capacity region
of the Gaussian MIMO degraded compound multi-receiver wiretap
channel when $K_2=1$ as follows.

\begin{Theo}
\label{theorem_mimo} The secrecy capacity region of the Gaussian
MIMO degraded compound channel when $K_2=1$ is given by the union
of rate pairs $(R_1,R_2)$ satisfying
\begin{align}
R_1 & \leq \min_{j=1,\ldots,K_1}\frac{1}{2} \log
\frac{|\bbk+\bbsigma_j^1|}{|\bbsigma_j^1|}-\frac{1}{2} \log
\frac{|\bbk+\bbsigma_Z|}{|\bbsigma_Z|} \\
R_2 & \leq \frac{1}{2} \log
\frac{|\bbs+\bbsigma^2|}{|\bbk+\bbsigma^2|} -\frac{1}{2} \log
\frac{|\bbs+\bbsigma_Z|}{|\bbk+\bbsigma_Z|}
\end{align}
where we dropped the subscript of $\bbsigma_k^2$ since $K_2=1$,
and the union is over all positive semi-definite matrices $\bbk$
such that $\bbk\preceq \bbs$.
\end{Theo}
The proof of this theorem is given in
Appendix~\ref{proof_of_theorem_mimo}.

We now consider the case $K_2>1$. We first note that since the
secrecy capacity region given in Theorem~\ref{theorem_discrete} is
convex, the boundary of this region can be written as the solution
of the following optimization problem
\begin{align}
\max_{(U,\bbx)}~ \min_{j=1,\ldots,K_1}R_{1j}+\mu
\min_{k=1,\ldots,K_2}R_{2k} \label{original_optimization}
\end{align}
where $R_{1j}$ and $R_{2k}$ are given by
\begin{align}
R_{1j}&=I(\bbx;\bby_{j}^1|U,\bbz)=I(\bbx;\bby_{j}^1|U)-I(\bbx;\bbz|U) \\
R_{2k}&=I(U;\bby_{k}^2|\bbz)=I(U;\bby_{k}^2)-I(U;\bbz)
\end{align}
respectively, and the maximization is over all $(U,\bbx)$ such
that $E\left[\bbx\bbx^{\top}\right]\preceq \bbs$. In the sequel,
we show that jointly Gaussian $(U,\bbx)$ is the maximizer for
(\ref{original_optimization}) when $\mu \leq 1$. To this end, we
need to consider the optimal Gaussian solution for
(\ref{original_optimization}), i.e., the solution of
(\ref{original_optimization}) when $(U,\bbx)$ is restricted to be
Gaussian. The corresponding optimization problem is
\begin{align}
\max_{\bzero\preceq \bbk\preceq \bbs}
~\min_{j=1,\ldots,K_1}R_{1j}^{G}(\bbk)+\mu
\min_{k=1,\ldots,K_2}R_{2k}^{G}(\bbk)
\label{gaussian_optimization}
\end{align}
where $R_{1j}^G(\bbk)$ and $R_{2k}^G(\bbk)$ are given by
\begin{align}
R_{1j}^G(\bbk)&= \frac{1}{2} \log
\frac{|\bbk+\bbsigma_j^1|}{|\bbsigma_j^1|}-\frac{1}{2}\log
\frac{|\bbk+\bbsigma_Z|}{|\bbsigma_Z|} \\
R_{2k}^G(\bbk)&= \frac{1}{2} \log
\frac{|\bbs+\bbsigma_k^2|}{|\bbk+\bbsigma_k^2|}-\frac{1}{2}\log
\frac{|\bbs+\bbsigma_Z|}{|\bbk+\bbsigma_Z|}
\end{align}
We assume that the maximum for (\ref{gaussian_optimization})
occurs at $\bbk=\bbk^*$, and the corresponding rate pair is
$(R_1^*,R_2^*)$\footnote{With this assumption, we implicitly
assume that the maximum in (\ref{gaussian_optimization}) occurs at
a single rate pair $(R_1^*,R_2^*)$. In fact, there might be more
than one rate pair where the maximum occurs. Even if this is the
case, we can simply consider only one of them, since our ultimate
goal is to show that the maximum in (\ref{original_optimization})
is equal to the maximum in (\ref{gaussian_optimization}).}, i.e.,
\begin{align}
R_1^*&=\min_{j=1,\ldots,K_1}R_{1j}^G(\bbk^*)\\
R_2^*&=\min_{k=1,\ldots,K_2}R_{2k}^G(\bbk^*)
\end{align}
The KKT conditions that this optimal covariance matrix $\bbk^*$
needs to satisfy are given in the following lemma.
\begin{Lem}
\label{lemma_gaussian_optimization} The optimal covariance matrix
for (\ref{gaussian_optimization}), $\bbk^*$, needs to satisfy
\begin{align}
\sum_{j=1}^{K_1}\lambda_{1j}
(\bbk^*+\bbsigma_j^1)^{-1}-(\bbk^*+\bbsigma_Z)^{-1}+\bbm=\mu
\sum_{k=1}^{K_2}\lambda_{2k} (\bbk^*+\bbsigma_k^2)^{-1}-\mu
(\bbk^*+\bbsigma_Z)^{-1}+\bbm_S \label{KKT_lemma}
\end{align}
where $\sum_{j=1}^{K_1}\lambda_{1j}=1$, and $\lambda_{1j}\geq 0$
with equality if $R_{1j}^G(\bbk^*)>R_1^*$;
$\sum_{k=1}^{K_2}\lambda_{2k}=1$, and $\lambda_{2k}\geq 0$ with
equality if $R_{2k}^G(\bbk^*)>R_2^*$; and $\bbm$ and $\bbm_S$ are
positive semi-definite matrices which satisfy
$\bbk^*\bbm=\bbm\bbk^*=\bzero$ and
$(\bbs-\bbk^*)\bbm_S=\bbm_S(\bbs-\bbk^*)=\bzero$, respectively.
\end{Lem}
The proof of this lemma is given
in~Appendix~\ref{proofs_for_partial_results}.

To show that both (\ref{original_optimization}) and
(\ref{gaussian_optimization}) have the same value when $\mu \leq
1$, we use the following optimization result due
to~\cite{Tie_Liu_Compound}.

\begin{Lem}[\!\!\cite{Tie_Liu_Compound}, Lemma~2]
\label{lemma_tie_liu} Let
$U,\bbx,\{\bbn_j^1\}_{j=1}^{K_1},\{\bbn_k^2\}_{k=1}^{K_2},\bbn_Z$
be as defined before. The following expression
\begin{align}
\sum_{j=1}^{K_1}\lambda_{1j}h(\bbx+\bbn_j^1|U)-\mu\sum_{k=1}^{K_2}\lambda_{2k}h(\bbx+\bbn_k^2|U)-(1-\mu)h(\bbx+\bbn_Z|U)
\end{align}
is maximized by jointly Gaussian $(U,\bbx)$ when $\mu\leq 1$.
Furthermore, the optimal covariance matrix needs to satisfy
(\ref{KKT_lemma}), where $\bbm$ and $\bbm_S$ are as they are
defined in Lemma~\ref{lemma_gaussian_optimization}.
\end{Lem}

In~\cite{Tie_Liu_Compound}, a weaker version of this lemma is
proved. This weaker version requires the existence of a covariance
matrix $\bbk^*$ for which the Lagrange multiplier $\bbm$ in
(\ref{KKT_lemma}) is zero. However, using the channel enhancement
technique~\cite{Shamai_MIMO}, this requirement can be removed.
Using Lemma~\ref{lemma_tie_liu} in conjunction with
Lemma~\ref{lemma_gaussian_optimization}, we are able to
characterize the secrecy capacity region partially for the case
$K_2> 1$.

\begin{Theo}
\label{theorem_partial} The boundary of the secrecy capacity
region of the degraded Gaussian MIMO compound multi-receiver
wiretap channel is given by the solution of the following
optimization problem
\begin{align}
\max_{\bzero\preceq \bbk\preceq \bbs} \min_{j=1,\ldots,K_1}
R_{1j}^G(\bbk)+\mu \min_{k=1,\ldots,K_2} R_{2k}^{G}(\bbk)
\end{align}
for $\mu\leq 1$. That is, for this part of the secrecy rate
region, jointly Gaussian auxiliary random variables and channel
inputs are optimal.
\end{Theo}
The proof of this theorem is given
in~Appendix~\ref{proofs_for_partial_results}.

\subsection{The Second Scenario: Layered Confidential Messages}

In the second scenario, the transmitter wants to send a
confidential message to users in the first group which needs to be
kept confidential from the second group of users and
eavesdroppers. The transmitter also wants to send a different
confidential message to users in the second group, which needs to
be kept confidential from the eavesdroppers. As opposed to the
first scenario, in this case, we do not put any restriction on the
number of eavesdroppers. The graphical illustration of the second
scenario is given in Figure~\ref{fig_compound_scenario_2}. The
situation where there is only one user in each group and one
eavesdropper was investigated in~\cite{Vector_Costa_EPI}. Hence,
this second scenario can be seen as a generalization of the model
in~\cite{Vector_Costa_EPI} to a compound channel setting.
Following the terminology of~\cite{Vector_Costa_EPI}, we call this
channel model the degraded compound multi-receiver wiretap channel
with {\it layered messages}.

An $(n,2^{nR_1},2^{nR_2})$ code for the degraded compound
multi-receiver wiretap channel with {\it layered messages}
consists of two message sets
$\mathcal{W}_1=\{1,\ldots,2^{nR_1}\},\mathcal{W}_2=\{1,\ldots,2^{nR_2}\}$
and an encoder $f:\mathcal{W}_1\times \mathcal{W}_2 \rightarrow
\mathcal{X}^n$, one decoder for each legitimate user in the first
group $g_{j}^1:\mathcal{Y}^{1,n}_j\rightarrow
\mathcal{W}_1,~j=1,\ldots,K_1$, and one decoder for each
legitimate user in the second group
$g_{k}^2:\mathcal{Y}^{2,n}_k\rightarrow
\mathcal{W}_2,~k=1,\ldots,K_2$. The probability of error is
defined as
\begin{align}
P_e^n=\max \{P_e^{1,n},P_{e}^{2,n}\}
\end{align}
where $P_e^{1,n}$ and $P_{e}^{2,n}$ are given by
\begin{align}
P_e^{1,n}=\max_{j\in\{1,\ldots,K_1\}} \Pr
\left[g_j^{1}(Y_{j}^{1,n})\neq W_1\right]\\
P_e^{2,n}=\max_{k\in\{1,\ldots,K_2\}} \Pr
\left[g_k^{2}(Y_{k}^{2,n})\neq W_2\right]
\end{align}
A secrecy rate pair is said to be achievable if there exists an
$(n,2^{nR_1},2^{nR_2})$ code which has
$\lim_{n\rightarrow\infty}P_e^n=0$,
\begin{align}
\lim_{n\rightarrow \infty} \frac{1}{n}I(W_2;Z_t^n)=0,\quad
t=1,\ldots,K_Z \label{perfect_secrecy_layered_1}
\end{align}
and
\begin{align}
\lim_{n\rightarrow \infty} \frac{1}{n}I(W_1;Y_k^{2,n}|W_2)=0,\quad
k=1,\ldots,K_2 \label{perfect_secrecy_layered_2}
\end{align}
We note that these two secrecy conditions imply
\begin{align}
\lim_{n\rightarrow \infty} \frac{1}{n}I(W_1,W_2;Z_t^n)=0,\quad
t=1,\ldots,K_Z
\end{align}
Furthermore, it is clear that we are only interested in perfect
secrecy rates of the channel. The secrecy capacity region is
defined as the closure of all achievable secrecy rate pairs. A
single-letter characterization of the secrecy capacity region is
given as follows.
\begin{Theo}
\label{theorem_discrete_layered} The secrecy capacity region of
the degraded compound multi-receiver wiretap channel with layered
messages is given by the union of rate pairs $(R_1,R_2)$
satisfying
\begin{align}
R_1 &\leq \min_{\substack{j=1,\ldots,K_1\\k=1,\ldots,K_2}}
I(X;Y^{1}_j|U,Y^2_k)\\
R_2 &\leq \min_{\substack{k=1,\ldots,K_2\\t=1,\ldots,K_Z}}
I(U;Y^{2}_k|Z_t)
\end{align}
where the union is over all random variable pairs $(U,X)$ such
that
\begin{align}
U\rightarrow X \rightarrow Y_j^1 \rightarrow Y^* \rightarrow
Y_{k}^2 \rightarrow Z^* \rightarrow Z_t
\end{align}
for any triple $(j,k,t)$.
\end{Theo}
The proof of this theorem is given in
Appendix~\ref{proof_of_theorem_discrete_layered}. Similar to the
converse proof of Theorem~\ref{theorem_discrete}, the presence of
the fictitious users $Y^*$ and $Z^*$ plays an important role here
as well. In particular, these two random variables introduce a
conditional independence structure to the channel which enables us
to define the auxiliary random variable $U$ that yields a tight
outer bound. Despite this similarity in the role of fictitious
users in converse proofs, there is a significant difference
between
Theorems~\ref{theorem_discrete}~and~\ref{theorem_discrete_layered};
in particular, it does not seem to be possible to extend
Theorem~\ref{theorem_discrete} to an arbitrary number of
eavesdroppers, while Theorem~\ref{theorem_discrete_layered} holds
for any number of eavesdroppers. This is due to the difference of
two communication scenarios. In the second scenario, since we
assume that users in the second group as well as the eavesdroppers
wiretap users in the first group, we are able to provide a
converse proof for the general situation of arbitrary number of
eavesdroppers.

As an aside, if we set $K_1=K_2=K_Z=1$, then as the degraded
compound multi-receiver wiretap channel with layered messages
reduces to the degraded multi-receiver wiretap channel with
layered messages of~\cite{Vector_Costa_EPI}, the secrecy capacity
region in Theorem~\ref{theorem_discrete_layered} reduces to the
secrecy capacity region of the channel model
in~\cite{Vector_Costa_EPI}.

\subsubsection{Parallel Degraded Compound Multi-receiver Wiretap Channels with Layered Messages}

In the next section, we investigate the Gaussian parallel degraded
compound multi-receiver wiretap channel with layered messages. To
that end, here we obtain the secrecy capacity region of the
parallel degraded compound multi-receiver wiretap channel with
layered messages in a single-letter form as follows.
\begin{Theo}
\label{theorem_discrete_layered_parallel} The secrecy capacity
region of the parallel degraded compound multi-receiver wiretap
channel with layered messages is given by the union of rate pairs
$(R_1,R_2)$ satisfying
\begin{align}
R_1&\leq \min_{\substack{j=1,\ldots,K_1 \\ k=1,\ldots,K_2}}
\sum_{\ell=1}^L I(X_{\ell};Y_{j\ell}^1|U_{\ell},Y_{k\ell}^2) \label{theorem_discrete_layered_parallel_R1} \\
R_2&\leq \min_{\substack{k=1,\ldots,K_2 \\ t=1,\ldots,K_Z}}
\sum_{\ell=1}^L I(U_{\ell};Y_{k\ell}^2|Z_{t\ell})
\label{theorem_discrete_layered_parallel_R2}
\end{align}
where the union is over all $\prod_{\ell=1}^L
p(u_{\ell},x_{\ell})$ such that
\begin{align}
U_{\ell} \rightarrow X_{\ell} \rightarrow Y_{j\ell}^1 \rightarrow
Y_{\ell}^*\rightarrow Y_{k\ell}^2 \rightarrow
Z_{\ell}^*\rightarrow Z_{t\ell}
\end{align}
for any $(\ell,j,k,t)$.
\end{Theo}

Since parallel degraded compound multi-receiver wiretap channels
with layered messages is a special case of the degraded compound
multi-receiver wiretap channel,
Theorem~\ref{theorem_discrete_layered} implicitly gives the
secrecy capacity region of parallel degraded compound
multi-receiver wiretap channels with layered messages. However, we
still need to show that the region in
Theorem~\ref{theorem_discrete_layered} is equivalent to the region
in Theorem~\ref{theorem_discrete_layered_parallel}. That is, we
need to prove the optimality of independent signalling in each
sub-channel. The proof of
Theorem~\ref{theorem_discrete_layered_parallel} is provided in
Appendix~\ref{proof_of_theorem_discrete_layered_parallel}.

\subsubsection{Gaussian Parallel Degraded Compound Multi-receiver Wiretap Channels with Layered Messages}

We now obtain the secrecy capacity region of Gaussian parallel
degraded compound multi-receiver wiretap channels with layered
messages. To that end, we need to evaluate the region given in
Theorem~\ref{theorem_discrete_layered_parallel}, i.e., we need to
find the optimal distribution $\prod_{\ell=1}^L
p(u_{\ell},x_{\ell})$. We first introduce the following theorem,
which is an extension of
Theorem~\ref{theorem_parallel_Gauss_optimization}.
\begin{Theo}
\label{theorem_parallel_layered_Gauss_optimization} Let
$N_1,N^*,N_2,\tilde{N},N_Z$ be zero-mean Gaussian random variables
with variances
$\sigma_1^2,\sigma_*^{2},\sigma_2^2,\tilde{\sigma}^2,\sigma_Z^2$,
respectively, where
\begin{align}
\sigma_1^2\leq \sigma_*^2 \leq \sigma_2^2 \leq \tilde{\sigma}^2
\leq \sigma_Z^2 \label{dummy_order_2}
\end{align}
Let $(U,X)$ be an arbitrarily dependent random variable pair,
which is independent of\break$(N_1,N^*,N_2,\tilde{N},N_Z)$, and
the second moment of $X$ be constrained as $E\left[X^2\right] \leq
P$. Then, for any feasible $(U,X)$, we can find a $P^*\leq P$ such
that
\begin{align}
h(X+\tilde{N}|U)-h(X+N^*|U)=\frac{1}{2} \log
\frac{P^*+\tilde{\sigma}^2}{P^*+\sigma_*^2}
\end{align}
and
\begin{align}
h(X+N_Z|U)-h(X+N_2|U) &\leq \frac{1}{2} \log
\frac{P^*+\sigma_Z^2}{P^*+\sigma_2^2}\label{parallel_layered_opt_1}\\
h(X+N_2|U)-h(X+N_1|U) &\geq \frac{1}{2} \log
\frac{P^*+\sigma_2^2}{P^*+\sigma_1^2}
\label{parallel_layered_opt_2}
\end{align}
for any $(\sigma_1^2,\sigma_2^2,\sigma_Z^2)$ satisfying the order
in (\ref{dummy_order_2}).
\end{Theo}
The proof of this theorem is given in
Appendix~\ref{proof_of_theorem_parallel_layered_Gauss_optimization}.
The proof of this theorem basically relies on
Theorem~\ref{theorem_parallel_Gauss_optimization} and Costa's
entropy power inequality~\cite{Costa_EPI}.

Using this theorem, we can establish the secrecy capacity region
of the Gaussian parallel degraded compound multi-receiver wiretap
channel with layered messages as follows.
\begin{Theo}
\label{theorem_parallel_Gaussian_layered} The secrecy capacity
region of the Gaussian parallel degraded compound multi-receiver
wiretap channel with layered messages is given by the union of
rate pairs $(R_1,R_2)$ satisfying
\begin{align}
R_1&\leq \min_{\substack{j=1,\ldots,K_1\\ k=1,\ldots,K_2}}
\sum_{\ell=1}^L \frac{1}{2} \log
\left(1+\frac{\beta_{\ell}P_{\ell}}{\Lambda_{j,\ell\ell}^1}\right)
-\frac{1}{2} \log
\left(1+\frac{\beta_{\ell}P_{\ell}}{\Lambda_{k,\ell\ell}^2}\right)\label{parallel_layered_R1}\\
R_2&\leq \min_{\substack{k=1,\ldots,K_2\\t=1,\ldots,K_Z}}
\sum_{\ell=1}^L \frac{1}{2} \log
\left(1+\frac{\bar{\beta}_{\ell}P_{\ell}}{\beta_{\ell}P_{\ell}+\Lambda_{k,\ell\ell}^2}\right)-\frac{1}{2}
\log
\left(1+\frac{\bar{\beta}_{\ell}P_{\ell}}{\beta_{\ell}P_{\ell}+\Lambda_{t,\ell\ell}^Z}\right)
\label{parallel_layered_R2}
\end{align}
where
$\bar{\beta}_{\ell}=1-\beta_{\ell}\in[0,1],~\ell=1,\ldots,L,$ and
the union is over all $\{P_{\ell}\}_{\ell=1}^L$ such that
$\sum_{\ell=1}^L P_{\ell}=P$.
\end{Theo}
The proof of this theorem is given in
Appendix~\ref{proof_of_theorem_parallel_Gaussian_layered}. Similar
to Theorem~\ref{theorem_parallel_Gaussian}, here also, $P_{\ell}$
denotes the amount of power $P$ devoted to the transmission in the
$\ell$th sub-channel. Similarly, $\beta_{\ell}$ is the fraction of
the power $P_{\ell}$ of the $\ell$th sub-channel spent for the
transmission to users in the first group.

\subsubsection{Gaussian MIMO Degraded Compound Multi-receiver Wiretap Channels with Layered Messages}

We now obtain the secrecy capacity region of the Gaussian MIMO
degraded compound multi-receiver wiretap channel with layered
messages. To that end, we need to evaluate the region given in
Theorem~\ref{theorem_discrete_layered}, i.e., find the optimal
random vector pair $(U,\bbx)$. We are able to find the optimal
random vector pair $(U,\bbx)$ when there is only one user in the
second group, i.e., $K_2=1$. To obtain that result, we first need
the following generalization of
Theorem~\ref{theorem_mimo_optimization}.
\begin{Theo}
\label{theorem_mimo_layered_optimization} Let
$(\bbn_1,\bbn_2,\bbn^*,\bbn_Z)$ be Gaussian random vectors with
covariance matrices $\bbsigma_1,\bbsigma_2,\bbsigma^*,\bbsigma_Z,$
respectively, where
\begin{align}
\bbsigma_1 \preceq \bbsigma_2 \preceq \bbsigma^* \preceq
\bbsigma_Z \label{dummy_order_3}
\end{align}
Let $(U,\bbx)$ be an arbitrarily dependent random vector pair,
which is independent of \break $(\bbn_1,\bbn_2,\bbn^*,\bbn_Z)$,
and the second moment of $\bbx$ be constrained as
$E\left[\bbx\bbx^\top\right] \preceq \bbs$. Then, for any feasible
$(U,\bbx)$, there exists a positive semi-definite matrix $\bbk^*$
such that $\bbk^*\preceq \bbs$, and it satisfies
\begin{align}
h(\bbx+\bbn^*|U)-h(\bbx+\bbn_2|U)=\frac{1}{2} \log
\frac{|\bbk^*+\bbsigma^*|}{|\bbk^*+\bbsigma_2|}
\end{align}
and
\begin{align}
h(\bbx+\bbn_Z|U)-h(\bbx+\bbn_2|U)&\leq \frac{1}{2} \log
\frac{|\bbk^*+\bbsigma_Z|}{|\bbk^*+\bbsigma_2|} \label{mimo_layered_inequality_1}\\
h(\bbx+\bbn_2|U)-h(\bbx+\bbn_1|U)&\geq \frac{1}{2} \log
\frac{|\bbk^*+\bbsigma_2|}{|\bbk^*+\bbsigma_1|}
\label{mimo_layered_inequality_2}
\end{align}
for any $(\bbsigma_1,\bbsigma_Z)$ satisfying the order in
(\ref{dummy_order_3}).
\end{Theo}
The proof of this theorem is given in
Appendix~\ref{proof_of_theorem_mimo_layered_optimization}. Using
this theorem, we can find the secrecy capacity region of the
Gaussian MIMO degraded compound multi-receiver wiretap channel
with layered messages when $K_2=1$ as follows.
\begin{Theo}
\label{theorem_mimo_layered} The secrecy capacity region of the
Gaussian MIMO degraded compound multi-receiver wiretap channel
with layered messages when $K_2=1$ is given by the union of rate
pairs $(R_1,R_2)$ satisfying
\begin{align}
R_1 &\leq \min_{j=1,\ldots,K_1} \frac{1}{2} \log
\frac{|\bbk+\bbsigma_j^1|}{|\bbsigma_j^1|}- \frac{1}{2} \log
\frac{|\bbk+\bbsigma^2|}{|\bbsigma^2|}\label{mimo_layered_R1}\\
R_2&\leq \min_{t=1,\ldots,K_Z} \frac{1}{2} \log
\frac{|\bbs+\bbsigma^2|}{|\bbk+\bbsigma^2|}-\frac{1}{2} \log
\frac{|\bbs+\bbsigma_t^Z|}{|\bbk+\bbsigma_t^Z|}
\label{mimo_layered_R2}
\end{align}
where the union is over all positive semi-definite matrices $\bbk$
such that $\bbk\preceq \bbs$.
\end{Theo}
The proof of this theorem is given in
Appendix~\ref{proof_of_theorem_mimo_layered}. As an aside, if we
set $K_1=K_Z=1$ in this theorem, we can recover the secrecy
capacity region of the degraded multi-receiver wiretap channel
with layered messages that was established
in~\cite{Vector_Costa_EPI}.

\section{Conclusions}

In this paper, we studied two different communication scenarios
for the degraded compound multi-receiver wiretap channel. In the
first scenario, the transmitter wants to send a confidential
message to users in the first group, and a different confidential
message to users in the second group, where both messages are to
be kept confidential from an eavesdropper. We establish the
secrecy capacity region of the general discrete memoryless channel
model, the parallel channel model, and the Gaussian parallel
channel model. For the Gaussian MIMO channel model, we obtain the
secrecy capacity region when there is only one user in the second
group. We also provide a partial characterization of the secrecy
capacity region when there are an arbitrary number of users in the
second group.

In the second scenario we study, the transmitter sends a
confidential message to users in the first group which is
wiretapped by both users in the second group and eavesdroppers. In
addition to this message sent to the first group of users, the
transmitter sends a different message to users in the second group
which needs to be kept confidential only from the eavesdroppers.
In this case, we do not put any restriction on the number of
eavesdroppers. As in the first scenario, we establish the secrecy
capacity region for the general discrete memoryless channel model,
the parallel channel model, and the Gaussian parallel channel
model. For the Gaussian MIMO channel model, we obtain the secrecy
capacity region when there is only one user in the second group.

\appendix

\appendixpage

\section{Proof of Theorem~\ref{theorem_discrete}}

\label{proof_of_theorem_discrete}

Achievability is clear. We provide the converse proof. For an
arbitrary code achieving the secrecy rates $(R_1,R_2)$, there
exist $(\epsilon_{1,n},\epsilon_{2,n})$ and $\gamma_{n}$ which
vanish as $n\rightarrow\infty$ such that
\begin{align}
H(W_1|Y_{j}^{1,n})&\leq n\epsilon_{1,n},\quad j=1,\ldots,K_1\label{Fano_lemma_implies_1}\\
H(W_2|Y_{k}^{2,n})&\leq n\epsilon_{2,n},\quad k=1,\ldots,K_2 \label{Fano_lemma_implies_2}\\
I(W_1,W_2;Z^n) &\leq n\gamma_n \label{secrecy_condition_implies}
\end{align}
where (\ref{Fano_lemma_implies_1}) and
(\ref{Fano_lemma_implies_2}) are due to Fano's lemma, and
(\ref{secrecy_condition_implies}) is due to the perfect secrecy
requirement stated in (\ref{perfect_secrecy}).

We define the following auxiliary random variables
\begin{align}
U_i=W_2 Y^{*,i-1}Z_{i+1}^n,\quad i=1,\ldots,n
\end{align}
which satisfy the following Markov chain
\begin{align}
U_i\rightarrow X_i \rightarrow Y_{j,i}^1 \rightarrow Y^*_{i}
\rightarrow Y^{2}_{k,i} \rightarrow Z_i, \quad i=1,\ldots,n
\label{def_auxiliary}
\end{align}
for any $(j,k)$ pair. The Markov chain in (\ref{def_auxiliary}) is
a consequence of the fact that the channel is memoryless and
degraded.

We first bound the rate of the second message:
\begin{align}
nR_2&=H(W_2) \\
&\leq I(W_2;Y_{k}^{2,n})+n \epsilon_{2,n}
\label{Fano_lemma_implies_1_1}\\
&\leq I(W_2;Y_{k}^{2,n})-I(W_2;Z^{n})+n
(\epsilon_{2,n}+\gamma_{n}) \label{secrecy_condition_implies_1}\\
&= I(W_2;Y_{k}^{2,n}|Z^{n})+n (\epsilon_{2,n}+\gamma_{n})
\label{channel_is_degraded_1} \\
&=\sum_{i=1}^n I(W_2;Y_{k,i}^{2}|Y_{k}^{2,i-1},Z^{n})+n
(\epsilon_{2,n}+\gamma_{n}) \\
&=\sum_{i=1}^n I(W_2;Y_{k,i}^{2}|Y_{k}^{2,i-1},Z_{i+1}^{n},Z_i)+n
(\epsilon_{2,n}+\gamma_{n}) \label{channel_is_degraded_2} \\
&\leq \sum_{i=1}^n
I(Y_{k}^{2,i-1},Z_{i+1}^{n},W_2;Y_{k,i}^{2}|Z_i)+n
(\epsilon_{2,n}+\gamma_{n})\\
&\leq \sum_{i=1}^n
I(Y^{*,i-1},Y_{k}^{2,i-1},Z_{i+1}^{n},W_2;Y_{k,i}^{2}|Z_i)+n
(\epsilon_{2,n}+\gamma_{n}) \\
&= \sum_{i=1}^n I(Y^{*,i-1},Z_{i+1}^{n},W_2;Y_{k,i}^{2}|Z_i)+n
(\epsilon_{2,n}+\gamma_{n}) \label{channel_is_degraded_3}\\
&=\sum_{i=1}^n I(U_i;Y_{k,i}^2|Z_i)+n(\epsilon_{2,n}+\gamma_n)
\label{almost_there}
\end{align}
where (\ref{Fano_lemma_implies_1_1}) is due to
(\ref{Fano_lemma_implies_2}), (\ref{secrecy_condition_implies_1})
is a consequence of (\ref{secrecy_condition_implies}),
(\ref{channel_is_degraded_1}) comes from the Markov chain
\begin{align}
W_2\rightarrow Y_{k}^{2,n}\rightarrow Z^n,\quad k=1,\ldots,K_2
\end{align}
which is a consequence of the fact that the channel is degraded,
(\ref{channel_is_degraded_2}) comes from the Markov chain
\begin{align}
Z^{i-1}\rightarrow Y_{k}^{2,i-1}\rightarrow
(Y_{k,i}^2,Z_i^n,W_2),\quad  k=1,\ldots,K_2
\end{align}
which is due to the fact that the channel is degraded and
memoryless, and (\ref{channel_is_degraded_3}) is a consequence of
the Markov chain
\begin{align}
Y_{k}^{2,i-1} \rightarrow Y^{*,i-1}\rightarrow
(W_2,Z_i^n,Y_{k,i}^2),\quad k=1,\ldots,K_2
\end{align}
which is due to the Markov chain in (\ref{our_Markov_chain}) and
the fact that the channel is memoryless.

Next we bound the rate of the first message:
\begin{align}
nR_1&=H(W_1)\\
&=H(W_1|W_2)\\
&\leq I(W_1;Y_j^{1,n}|W_2)+n\epsilon_{1,n}\label{Fano_lemma_implies_2_1}\\
&\leq
I(W_1;Y_j^{1,n}|W_2)-I(W_1;Z^{n}|W_2)+n(\epsilon_{1,n}+\gamma_{n})\label{secrecy_condition_implies_2}\\
&= I(W_1;Y_j^{1,n}|W_2,Z^n)+n(\epsilon_{1,n}+\gamma_{n})
\label{channel_is_degraded_4}\\
&=\sum_{i=1}^n
I(W_1;Y_{j,i}^{1}|W_2,Z^n,Y_j^{1,i-1})+n(\epsilon_{1,n}+\gamma_{n})\\
&=\sum_{i=1}^n
I(W_1;Y_{j,i}^{1}|W_2,Z_{i+1}^n,Y_j^{1,i-1},Z_i)+n(\epsilon_{1,n}+\gamma_{n})
\label{channel_is_degraded_5}\\
&=\sum_{i=1}^n
I(W_1;Y_{j,i}^{1}|W_2,Z_{i+1}^n,Y_j^{1,i-1},Y^{*,i-1},Z_i)+n(\epsilon_{1,n}+\gamma_{n})
\label{channel_is_degraded_6}\\
&\leq \sum_{i=1}^n
I(X_i,W_1;Y_{j,i}^{1}|W_2,Z_{i+1}^n,Y_j^{1,i-1},Y^{*,i-1},Z_i)+n(\epsilon_{1,n}+\gamma_{n})\\
&= \sum_{i=1}^n
I(X_i;Y_{j,i}^{1}|W_2,Z_{i+1}^n,Y_j^{1,i-1},Y^{*,i-1},Z_i)+n(\epsilon_{1,n}+\gamma_{n})
\label{channel_memoryless}\\
&= \sum_{i=1}^n
H(Y_{j,i}^{1}|W_2,Z_{i+1}^n,Y_j^{1,i-1},Y^{*,i-1},Z_i)
-H(Y_{j,i}^{1}|W_2,Z_{i+1}^n,Y_j^{1,i-1},Y^{*,i-1},Z_i,X_i)\nonumber\\
&\quad +n(\epsilon_{1,n}+\gamma_{n})\\
&\leq \sum_{i=1}^n H(Y_{j,i}^{1}|W_2,Z_{i+1}^n,Y^{*,i-1},Z_i)
-H(Y_{j,i}^{1}|W_2,Z_{i+1}^n,Y_j^{1,i-1},Y^{*,i-1},Z_i,X_i)\nonumber\\
&\quad +n(\epsilon_{1,n}+\gamma_{n})\label{conditioning_cannot}\\
&= \sum_{i=1}^n H(Y_{j,i}^{1}|W_2,Z_{i+1}^n,Y^{*,i-1},Z_i)
-H(Y_{j,i}^{1}|W_2,Z_{i+1}^n,Y^{*,i-1},Z_i,X_i)
+n(\epsilon_{1,n}+\gamma_{n}) \label{channel_memoryless_1}\\
&= \sum_{i=1}^n I(X_i;Y_{j,i}^{1}|W_2,Z_{i+1}^n,Y^{*,i-1},Z_i)
+n(\epsilon_{1,n}+\gamma_{n})\\
&= \sum_{i=1}^n I(X_i;Y_{j,i}^{1}|U_i,Z_i)
+n(\epsilon_{1,n}+\gamma_{n}) \label{almost_there_1}
\end{align}
where (\ref{Fano_lemma_implies_2_1}) is due to
(\ref{Fano_lemma_implies_1}), (\ref{secrecy_condition_implies_2})
is a consequence of (\ref{secrecy_condition_implies}),
(\ref{channel_is_degraded_4}) comes from the Markov chain
\begin{align}
(W_2,W_1)\rightarrow Y_j^{1,n} \rightarrow Z^n,\quad
j=1,\ldots,K_1
\end{align}
which is due to the fact that the channel is degraded,
(\ref{channel_is_degraded_5}) comes from the Markov chain
\begin{align}
Z^{i-1}\rightarrow Y_{j}^{1,i-1} \rightarrow
(W_1,W_2,Y_{j,i}^{1},Z_i^n),\quad j=1,\ldots,K_1
\end{align}
which is a consequence of the fact that the channel is degraded
and memoryless, (\ref{channel_is_degraded_6}) follows from the
Markov chain
\begin{align}
Y^{*,i-1}\rightarrow Y_{j}^{1,i-1}\rightarrow
(W_1,W_2,Y_{j,i}^1,Z_{i}^n),\quad j=1,\ldots,K_1
\end{align}
which results from the Markov chain in (\ref{our_Markov_chain})
and the fact that the channel is memoryless,
(\ref{channel_memoryless}) is a consequence of the Markov chain
\begin{align}
(Y_{j,i}^1,Z_i) \rightarrow X_i \rightarrow
(Y^{*,i-1},Y_j^{1,i-1},Z_{i+1}^n,W_1,W_2),\quad j=1,\ldots,K_1
\label{channel_memoryless_implies}
\end{align}
which is due to the fact that the channel is memoryless,
(\ref{conditioning_cannot}) comes from the fact that conditioning
cannot increase entropy, and (\ref{channel_memoryless_1}) is again
due to the Markov chain in (\ref{channel_memoryless_implies}).

Next, we define a uniformly distributed random variable
$Q\in\{1,\ldots,n\}$, and $U=(Q,U_{Q}),X=X_{Q},Y_{j}^1=Y_{j,Q}^1,
Y_{k}^2=Y_{k,Q}^2, \textrm{ and } Z=Z_Q$. Using these definitions
in (\ref{almost_there}) and (\ref{almost_there_1}), we obtain the
single-letter expressions in Theorem~\ref{theorem_discrete}.

\section{Proof of Theorem~\ref{theorem_discrete_parallel}}

\label{proof_of_theorem_discrete_parallel}

The achievability of this region follows from
Theorem~\ref{theorem_discrete} by selecting
$(U,X)=(U_1,X_1,\ldots,U_L,\break X_L)$ with a joint distribution
of the product form $p(u,x)=\prod_{\ell=1}^L
p(u_{\ell},x_{\ell})$. We next provide the converse proof. To that
end, we define the following auxiliary random variables
\begin{align}
U_{\ell,i}=W_2 Y^{*,i-1} Z_{i+1}^n Y^*_{[1:\ell-1],i}
Z_{[\ell+1:L],i},\quad i=1,\ldots,n, \quad \ell=1,\ldots,L
\end{align}
which satisfy the Markov chain
\begin{align}
U_{\ell,i}\rightarrow X_{\ell,i} \rightarrow
(Y_{j\ell,i}^1,Y_{k\ell,i}^2,Z_{\ell,i})
\end{align}
for any $(j,k,\ell)$ triple because of the facts that the channel
is memoryless and sub-channels are independent.

We bound the rate of the second message. Following the same steps
as in the converse proof of Theorem~\ref{theorem_discrete}, we get
to (\ref{channel_is_degraded_2}). Then,
\begin{align}
nR_2& \leq \sum_{i=1}^n
I(W_2;Y_{k,i}^2|Y_{k}^{2,i-1},Z_{i+1}^n,Z_i)+n(\epsilon_{2,n}+\gamma_n)\\
&=\sum_{i=1}^n \sum_{\ell=1}^L
I(W_2;Y_{k\ell,i}^2|Y_{k}^{2,i-1},Z_{i+1}^n,Z_i,Y_{k[1:\ell-1],i}^2)+n(\epsilon_{2,n}+\gamma_n)\\
&= \sum_{i=1}^n \sum_{\ell=1}^L
I(W_2;Y_{k\ell,i}^2|Y_{k}^{2,i-1},Z_{i+1}^n,Z_{[\ell+1:L],i},Y_{k[1:\ell-1],i}^2,Z_{\ell,i})+n(\epsilon_{2,n}+\gamma_n)
\label{channel_is_degraded_7} \\
&\leq \sum_{i=1}^n \sum_{\ell=1}^L
I(Y_{k}^{2,i-1},Z_{i+1}^n,Z_{[\ell+1:L],i},Y_{k[1:\ell-1],i}^2,W_2;Y_{k\ell,i}^2|Z_{\ell,i})+n(\epsilon_{2,n}+\gamma_n)\\
&\leq \sum_{i=1}^n \sum_{\ell=1}^L
I(Y_{k}^{2,i-1},Y^{*,i-1},Z_{i+1}^n,Z_{[\ell+1:L],i},Y_{k[1:\ell-1],i}^2,Y^{*}_{[1:\ell-1],i},W_2;Y_{k\ell,i}^2|Z_{\ell,i})\nonumber\\
&\quad +n(\epsilon_{2,n}+\gamma_n)\\
&= \sum_{i=1}^n \sum_{\ell=1}^L
I(Y^{*,i-1},Z_{i+1}^n,Z_{[\ell+1:L],i},Y^{*}_{[1:\ell-1],i},W_2;Y_{k\ell,i}^2|Z_{\ell,i})
+n(\epsilon_{2,n}+\gamma_n) \label{channel_is_degraded_8} \\
&= \sum_{i=1}^n \sum_{\ell=1}^L
I(U_{\ell,i};Y_{k\ell,i}^2|Z_{\ell,i}) +n(\epsilon_{2,n}+\gamma_n)
\label{almost_there_2}
\end{align}
where (\ref{channel_is_degraded_7}) follows from the Markov chain
\begin{align}
Z_{[1:\ell-1],i}\rightarrow Y_{k[1:\ell-1],i}^2 \rightarrow (
W_2,Y_{k}^{2,i-1},Z_{i+1}^n,Z_{[\ell:L],i},Y_{k\ell,i}^2 )
\end{align}
which is a consequence of the facts that the channel is degraded
and memoryless, and sub-channels are independent, and
(\ref{channel_is_degraded_8}) is due to the Markov chain
\begin{align}
(Y_{k}^{2,i-1},Y_{k[1:\ell-1],i}^2 ) \rightarrow
(Y^{*,i-1},Y_{[1:\ell-1],i}^* )\rightarrow
(W_2,Z_{i+1}^n,Z_{[\ell:L],i},Y_{k\ell,i}^2)
\end{align}
which is a consequence of the Markov chain in
(\ref{our_Markov_chain_1}) and the facts that the channel is
memoryless and sub-channels are independent.

We next bound the rate of the first message. Again, following the
same steps as in the converse proof of
Theorem~\ref{theorem_discrete}, we get to
(\ref{channel_is_degraded_5}). Then,
\begin{align}
nR_1 & \leq \sum_{i=1}^n
I(W_1;Y_{j,i}^1|W_2,Y_{j}^{1,i-1},Z_{i+1}^n,Z_i)+n(\epsilon_{1,n}+\gamma_n)\\
&=\sum_{i=1}^n \sum_{\ell=1}^L
I(W_1;Y_{j\ell,i}^1|W_2,Y_{j}^{1,i-1},Z_{i+1}^n,Y_{j[1:\ell-1],i}^1,Z_i)+n(\epsilon_{1,n}+\gamma_n)\\
&=\sum_{i=1}^n \sum_{\ell=1}^L
I(W_1;Y_{j\ell,i}^1|W_2,Y_{j}^{1,i-1},Z_{i+1}^n,Y_{j[1:\ell-1],i}^1,Z_{[\ell+1:L],i},Z_{\ell,i})+n(\epsilon_{1,n}+\gamma_n)
\label{channel_is_degraded_9}\\
&=\sum_{i=1}^n \sum_{\ell=1}^L
I(W_1;Y_{j\ell,i}^1|W_2,Y_{j}^{1,i-1},Y^{*,i-1},Z_{i+1}^n,Y_{j[1:\ell-1],i}^1,Y^*_{[1:\ell-1],i},Z_{[\ell+1:L],i},Z_{\ell,i})\nonumber\\
&\quad +n(\epsilon_{1,n}+\gamma_n) \label{channel_is_degraded_10}
\\
&\leq \sum_{i=1}^n \sum_{\ell=1}^L
I(X_{\ell,i},W_1;Y_{j\ell,i}^1|W_2,Y_{j}^{1,i-1},Y^{*,i-1},Z_{i+1}^n,Y_{j[1:\ell-1],i}^1,Y^*_{[1:\ell-1],i},Z_{[\ell+1:L],i},Z_{\ell,i})\nonumber\\
&\quad +n(\epsilon_{1,n}+\gamma_n)\\
&= \sum_{i=1}^n \sum_{\ell=1}^L
I(X_{\ell,i};Y_{j\ell,i}^1|W_2,Y_{j}^{1,i-1},Y^{*,i-1},Z_{i+1}^n,Y_{j[1:\ell-1],i}^1,Y^*_{[1:\ell-1],i},Z_{[\ell+1:L],i},Z_{\ell,i})\nonumber\\
&\quad +n(\epsilon_{1,n}+\gamma_n) \label{channel_memoryless_2} \\
&= \sum_{i=1}^n \sum_{\ell=1}^L
H(Y_{j\ell,i}^1|W_2,Y_{j}^{1,i-1},Y^{*,i-1},Z_{i+1}^n,Y_{j[1:\ell-1],i}^1,Y^*_{[1:\ell-1],i},Z_{[\ell+1:L],i},Z_{\ell,i})\nonumber\\
&\quad
-H(Y_{j\ell,i}^1|W_2,Y_{j}^{1,i-1},Y^{*,i-1},Z_{i+1}^n,Y_{j[1:\ell-1],i}^1,Y^*_{[1:\ell-1],i},Z_{[\ell+1:L],i},Z_{\ell,i},X_{\ell,i})
\nonumber\\
&\quad +n(\epsilon_{1,n}+\gamma_n) \\
&\leq  \sum_{i=1}^n \sum_{\ell=1}^L
H(Y_{j\ell,i}^1|W_2,Y^{*,i-1},Z_{i+1}^n,Y^*_{[1:\ell-1],i},Z_{[\ell+1:L],i},Z_{\ell,i})\nonumber\\
&\quad
-H(Y_{j\ell,i}^1|W_2,Y_{j}^{1,i-1},Y^{*,i-1},Z_{i+1}^n,Y_{j[1:\ell-1],i}^1,Y^*_{[1:\ell-1],i},Z_{[\ell+1:L],i},Z_{\ell,i},X_{\ell,i})
\nonumber\\
&\quad +n(\epsilon_{1,n}+\gamma_n) \label{conditioning_cannot_1}\\
&= \sum_{i=1}^n \sum_{\ell=1}^L
H(Y_{j\ell,i}^1|W_2,Y^{*,i-1},Z_{i+1}^n,Y^*_{[1:\ell-1],i},Z_{[\ell+1:L],i},Z_{\ell,i})\nonumber\\
&\quad
-H(Y_{j\ell,i}^1|W_2,Y^{*,i-1},Z_{i+1}^n,Y^*_{[1:\ell-1],i},Z_{[\ell+1:L],i},Z_{\ell,i},X_{\ell,i})
+n(\epsilon_{1,n}+\gamma_n) \label{channel_memoryless_3} \\
&= \sum_{i=1}^n \sum_{\ell=1}^L
I(X_{\ell,i};Y_{j\ell,i}^1|W_2,Y^{*,i-1},Z_{i+1}^n,Y^*_{[1:\ell-1],i},Z_{[\ell+1:L],i},Z_{\ell,i})
+n(\epsilon_{1,n}+\gamma_n)\\
&=\sum_{i=1}^n \sum_{\ell=1}^L
I(X_{\ell,i};Y_{j\ell,i}^1|U_{\ell,i},Z_{\ell,i})+n(\epsilon_{1,n}+\gamma_n)
\label{almost_there_3}
\end{align}
where (\ref{channel_is_degraded_9}) follows from the Markov chain
\begin{align}
Z_{[1:\ell-1],i}\rightarrow Y_{j[1:\ell-1],i}^1\rightarrow
(W_1,W_2,Y_{j}^{1,i-1},Z_{i+1}^n,Y_{j\ell,i}^1,Z_{[\ell:L],i})
\end{align}
which is due to the facts that the channel is degraded and
memoryless, and sub-channels are independent,
(\ref{channel_is_degraded_10}) comes from the Markov chain
\begin{align}
(Y^{*,i-1},Y^*_{[1:\ell-1],i})\rightarrow
(Y_{j}^{1,i-1},Y_{j[1:\ell-1],i}^{1}) \rightarrow
(W_1,W_2,Z_{i+1}^n,Z_{[\ell:L],i},Y_{j\ell,i}^1)
\end{align}
which results from the Markov chain in (\ref{our_Markov_chain_1})
and the facts that the channel is memoryless, and sub-channels are
independent, (\ref{channel_memoryless_2}) comes from the Markov
chain
\begin{align}
(Y_{j\ell,i}^1,Z_{\ell,i})\rightarrow X_{\ell,i} \rightarrow
(W_1,W_2,Y_{j}^{1,i-1},Y^{*,i-1},Z_{i+1}^n,Y_{j[1:\ell-1],i}^1,Y^*_{[1:\ell-1],i},Z_{[\ell+1:L],i})
\label{markov_chain_ch_mless}
\end{align}
which is a consequence of the facts that the channel is
memoryless, and sub-channels are independent,
(\ref{conditioning_cannot_1}) results from the fact that
conditioning cannot increase entropy, and
(\ref{channel_memoryless_3}) is due to the Markov chain in
(\ref{markov_chain_ch_mless}).

Next, we a define a uniformly distributed random variable
$Q\in\{1,\ldots,n\}$,  and
$U_{\ell}=(Q,U_{\ell,Q}),X=X_{\ell,Q},Y_{j\ell}^1=Y_{j\ell,Q}^1,
Y_{k\ell}^2=Y_{k\ell,Q}^2, \textrm{ and } Z_\ell=Z_{\ell,Q}$.
Using these definitions in (\ref{almost_there_2}) and
(\ref{almost_there_3}), we obtain the single-letter expressions in
 Theorem~\ref{theorem_discrete_parallel}. Finally, we note that although auxiliary random variables
$\{U_\ell\}_{\ell=1}^L$ are dependent, their joint distribution
does not affect the bounds in
Theorem~\ref{theorem_discrete_parallel}. Thus, without loss of
generality, we can select them to be independent.

\section{Proof of Theorem~\ref{theorem_parallel_Gauss_optimization}}

\label{proof_of_theorem_parallel_Gauss_optimization}

We first note that
\begin{align}
\frac{1}{2} \log \frac{\sigma_*^2}{\sigma_Z^2} \leq
h(X+N^*|U)-h(X+N_Z|U)\leq \frac{1}{2} \log
\frac{P+\sigma_*^2}{P+\sigma_Z^2} \label{dummy_1}
\end{align}
where the right-hand side can be shown via the entropy power
inequality~\cite{Stam,Blachman}. To show the left-hand side, let
us define a Gaussian random variable $\tilde{N}$ with variance
$\sigma_Z^2-\sigma_*^2$, and independent of $(U,X,N^*)$. Thus, we
can write down the difference of differential entropy terms in
(\ref{dummy_1}) as
\begin{align}
h(X+N^*|U)-h(X+N_Z|U)&=h(X+N^*|U)-h(X+N^*+\tilde{N}|U)\\
&=-I(\tilde{N};X+N^*+\tilde{N}|U)\\
&=-h(\tilde{N}|U)+h(\tilde{N}|U,X+N^*+\tilde{N})\\
&\geq -h(\tilde{N}|U)+h(\tilde{N}|U,X+N^*+\tilde{N},X)
\label{conditioning_cannot_2}\\
&= -h(\tilde{N})+h(\tilde{N}|N^*+\tilde{N}) \label{independence_etc}\\
&=\frac{1}{2}\log \frac{\sigma_*^2}{\sigma_Z^2}
\end{align}
where (\ref{conditioning_cannot_2}) is due to the fact that
conditioning cannot increase entropy and (\ref{independence_etc})
is a consequence of the fact that $(U,X)$ and $(N^*,\tilde{N})$
are independent.

Equation (\ref{dummy_1}) implies that there exists $P^*$ such that
$P^*\leq P$ and
\begin{align}
h(X+N^*|U)-h(X+N_Z|U)= \frac{1}{2} \log
\frac{P^*+\sigma_*^2}{P^*+\sigma_Z^2} \label{fix_it}
\end{align}
which will be used frequently hereafter.

We now state Costa's entropy power inequality~\cite{Costa_EPI}
which will be used in the upcoming proof\footnote{Although,
Theorem~1 of \cite{Costa_EPI} states the inequality for a constant
$U$, using Jensen's inequality, the current form of the inequality
for an arbitrary $U$ can be shown.}.

\begin{Lem}[\!\!\cite{Costa_EPI}, Theorem~1] Let $(U,X)$ be an arbitrarily dependent
random variable pair, which is independent of $N$, where $N$ is a
Gaussian random variable. Then, we have
\begin{align}
e^{2h(X+\sqrt{t}N|U)} \geq (1-t)e^{2h(X|U)}+te^{2h(X+N|U)},\quad
0\leq t\leq 1
\end{align}
\end{Lem}

We now consider (\ref{inequality_1}). We first note that we can
write $N^*$ as
\begin{align}
N^*=N_1+\sqrt{t_1}\tilde{N}_1 \label{stability_2}
\end{align}
where $\tilde{N}_1$ is a Gaussian random variable with variance
$\sigma_Z^2-\sigma_1^2$, which is independent of $(U,X,N_1)$.
$t_1$ in (\ref{stability_2}) is given by
\begin{align}
t_1=\frac{\sigma_*^2-\sigma_1^2}{\sigma_Z^2-\sigma_1^2}
\label{t_2}
\end{align}
where it is clear that $t_1\in[0,1]$. Using (\ref{stability_2})
and Costa's entropy power inequality~\cite{Costa_EPI}, we get
\begin{align}
e^{2h(X+N^*|U)}&=e^{2h(X+N_1+\sqrt{t_1}\tilde{N}_1|U)}\\
&\geq (1-t_1) e^{2h(X+N_1|U)}+t_1 e^{2h(X+N_Z|U)}
\end{align}
which is equivalent to
\begin{align}
(1-t_1) e^{2\left[h(X+N_1|U)-h(X+N_Z|U)\right]}+t_1  &\leq
e^{2\left[h(X+N^*|U)-h(X+N_Z|U)\right]}\\
&= \frac{P^*+\sigma_*^2}{P^*+\sigma_Z^2} \label{dummy_2}
\end{align}
where (\ref{dummy_2}) is obtained by using (\ref{fix_it}).
Equation (\ref{dummy_2}) is equivalent to
\begin{align}
h(X+N_1|U)-h(X+N_Z|U) &\leq \frac{1}{2}\log
\frac{1}{1-t_1}\left(\frac{P^*+\sigma_*^2}{P^*+\sigma_Z^2}-t_1\right)\\
&= \frac{1}{2}\log
\left(\frac{P^*}{P^*+\sigma_Z^2}+\frac{1}{1-t_1}\frac{\sigma_*^2-t_1\sigma_Z^2}{P^*+\sigma_Z^2}\right)\\
&= \frac{1}{2}\log \frac{P^*+\sigma_1^2}{P^*+\sigma_Z^2}
\label{def_t_2_implies}
\end{align}
where we used the definition of $t_1$ given in (\ref{t_2}) to
obtain (\ref{def_t_2_implies}). Equation (\ref{def_t_2_implies})
proves (\ref{inequality_1}).

We now consider (\ref{inequality_2}). First, we note that we can
write $N_2$
\begin{align}
N_2=N^*+\sqrt{t_2}\tilde{N}_Z \label{stability_1}
\end{align}
where $\tilde{N}_Z$ is a Gaussian random variable with variance
$\sigma_Z^2-\sigma_*^2$, which is independent of $(U,X,N^*)$.
$t_2$ in (\ref{stability_1}) is given by
\begin{align}
t_2=\frac{\sigma_2^2-\sigma_*^2}{\sigma_Z^2-\sigma_*^2}
\label{t_1}
\end{align}
where it is clear that $t_2\in[0,1]$. Using (\ref{stability_1})
and Costa's entropy power inequality~\cite{Costa_EPI}, we get
\begin{align}
e^{2h(X+N_2|U)}&=e^{2h(X+N^*+\sqrt{t_2}\tilde{N}_Z|U)}\\
&\geq (1-t_2)e^{2h(X+N^*|U)}+t_2 e^{2h(X+N_Z|U)}
\end{align}
which is equivalent to
\begin{align}
e^{2\left[h(X+N_2|U)-h(X+N_Z|U)\right]} &\geq
(1-t_2)e^{2\left[h(X+N^*|U)-h(X+N_Z|U)\right]}+t_2 \\
&= (1-t_2)\frac{P^*+\sigma_*^2}{P^*+\sigma_Z^2}+t_2\\
&=\frac{P^*+\sigma_2^2}{P^*+\sigma_Z^2} \label{def_t_1_implies}
\end{align}
where (\ref{def_t_1_implies}) is obtained by using the definition
of $t_2$ given in (\ref{t_1}). Equation (\ref{def_t_1_implies}) is
equivalent to
\begin{align}
h(X+N_Z|U)-h(X+N_2|U) \leq
\frac{1}{2}\log\frac{P^*+\sigma_Z^2}{P^*+\sigma_2^2}
\end{align}
which is (\ref{inequality_2}). This completes the proof of
Theorem~\ref{theorem_parallel_Gauss_optimization}.

\section{Proof of Theorem~\ref{theorem_parallel_Gaussian}}

\label{proof_of_theorem_parallel_Gaussian}

Achievability is clear. We provide the converse proof. To this
end, let us fix the distribution $\prod_{\ell=1}^L
p(u_{\ell},x_{\ell})$ such that
\begin{align}
E\left[X_{\ell}^2\right]=P_{\ell},~\quad \ell=1,\ldots,L
\end{align}
and $\sum_{\ell=1}^L P_{\ell}\leq P$. We first establish the bound
on $R_2$ given in (\ref{parallel_Gaussian_R2}). To this end, we
start with (\ref{theorem_discrete_parallel_R2}). Using the Markov
chain $U_{\ell}\rightarrow Y_{k\ell}^2 \rightarrow Z_{\ell}$, we
have
\begin{align}
R_2 & \leq \min_{k=1,\ldots,K_2} \sum_{\ell=1}^L
I(U_{\ell};Y_{k\ell}^2)-I(U_{\ell};Z_{\ell})\label{theorem_discrete_parallel_implies}\\
&= \min_{k=1,\ldots,K_2} \sum_{\ell=1}^L
\left[h(Y_{k\ell}^2)-h(Z_{\ell})\right]+
\left[h(Z_{\ell}|U)-h(Y_{k\ell}^2|U)\right]\\
&\leq \min_{k=1,\ldots,K_2} \sum_{\ell=1}^L \frac{1}{2} \log
\frac{P_{\ell}+\Lambda^2_{k,\ell\ell}}{P_{\ell}+\Lambda_{Z,\ell\ell}}
+ \left[h(Z_{\ell}|U)-h(Y_{k\ell}^2|U)\right] \label{epi_implies}
\end{align}
where (\ref{epi_implies}) comes from the fact that Gaussian
$X_{\ell}$ maximizes
\begin{align}
h(Y_{k\ell}^2)-h(Z_{\ell})
\end{align}
which can be shown via the entropy power
inequality~\cite{Stam,Blachman}. We now use
Theorem~\ref{theorem_parallel_Gauss_optimization}. For that
purpose, we introduce the diagonal covariance matrix $\bblambda^*$
which satisfies
\begin{align}
\bblambda_{j}^1 \preceq \bblambda^{*} \preceq \bblambda_{k}^2
\end{align}
for any $(j,k)$ pair, and in particular, for the diagonal elements
of these matrices, we have
\begin{align}
\Lambda_{j,\ell\ell}^1 \leq \Lambda_{\ell\ell}^{*} \leq
\Lambda_{k,\ell\ell}^2
\end{align}
for any triple $(j,k,\ell)$. Thus, due to
Theorem~\ref{theorem_parallel_Gauss_optimization}, for any
selection of $\{(U_{\ell},X_{\ell})\}_{\ell=1}^L$, there exists a
$P_{\ell}^*$ such that
\begin{align}
P_{\ell}^* &\leq P_{\ell} \label{theorem_parallel_Gauss_optimization_implies_0}\\
h(Z_{\ell}|U_{\ell})-h(Y^1_{j\ell}|U_{\ell}) &\geq \frac{1}{2}
\log
\frac{P_{\ell}^*+\Lambda_{Z,\ell\ell}}{P_{\ell}^*+\Lambda^1_{j,\ell\ell}}
\label{theorem_parallel_Gauss_optimization_implies_1}\\
h(Z_{\ell}|U_{\ell})-h(Y^2_{k\ell}|U_{\ell}) &\leq \frac{1}{2}
\log
\frac{P_{\ell}^*+\Lambda_{Z,\ell\ell}}{P_{\ell}^*+\Lambda^2_{k,\ell\ell}}
\label{theorem_parallel_Gauss_optimization_implies_2}
\end{align}
for any triple $(j,k,\ell)$. Using
(\ref{theorem_parallel_Gauss_optimization_implies_2}) in
(\ref{epi_implies}), we get
\begin{align}
R_2 &\leq   \min_{k=1,\ldots,K_2} \sum_{\ell=1}^L \frac{1}{2} \log
\frac{P_{\ell}+\Lambda^2_{k,\ell\ell}}{P_{\ell}^*+\Lambda^2_{k,\ell\ell}}
- \frac{1}{2} \log
\frac{P_{\ell}+\Lambda_{Z,\ell\ell}}{P_{\ell}^*+\Lambda_{Z,\ell\ell}}
\end{align}
We define $P^*_{\ell}=\beta_{\ell}P_{\ell}$ and
$\bar{\beta}_{\ell}=1-\beta_{\ell},~\ell=1,\ldots,L$, where
$\beta_{\ell}\in[0,1]$ due to
(\ref{theorem_parallel_Gauss_optimization_implies_0}). Thus, we
have established the desired bound on $R_2$ given in
(\ref{parallel_Gaussian_R2}). We now bound $R_1$. We start with
(\ref{theorem_discrete_parallel_R1}). Using the Markov chain
$(U_{\ell},X_{\ell})\rightarrow Y_{j\ell}^1 \rightarrow Z_{\ell}$,
we have
\begin{align}
R_1 &\leq \min_{j=1,\ldots,K_1} \sum_{\ell=1}^L
I(X_{\ell};Y_{j\ell}^1|U_{\ell}) -I(X_{\ell};Z_{\ell}|U_{\ell})
\label{theorem_discrete_parallel_implies_1}\\
&= \min_{j=1,\ldots,K_1} \sum_{\ell=1}^L h(Y_{j\ell}^1|U_{\ell})
-h(Z_{\ell}|U_{\ell})- \frac{1}{2} \log
\frac{\Lambda_{j,\ell\ell}^1}{\Lambda_{Z,\ell\ell}}\\
&\leq \min_{j=1,\ldots,K_1} \sum_{\ell=1}^L \frac{1}{2} \log
\frac{P_{\ell}^*+\Lambda^1_{j,\ell\ell}}{P_{\ell}^*+\Lambda_{Z,\ell\ell}}
- \frac{1}{2} \log
\frac{\Lambda_{j,\ell\ell}^1}{\Lambda_{Z,\ell\ell}}
\label{theorem_parallel_Gauss_optimization_implies_3}
\end{align}
where (\ref{theorem_parallel_Gauss_optimization_implies_3}) comes
from (\ref{theorem_parallel_Gauss_optimization_implies_1}). Since
we defined $P_{\ell}^*=\beta_{\ell}P_{\ell}$,
(\ref{theorem_parallel_Gauss_optimization_implies_3}) is the
desired bound on $R_1$ given in (\ref{parallel_Gaussian_R1}),
completing the proof.

\section{Proof of Theorem~\ref{theorem_mimo}}

\label{proof_of_theorem_mimo}

The main tools for the proof of Theorem~\ref{theorem_mimo} are
Theorem~\ref{theorem_mimo_optimization}, and the following
so-called worst additive noise lemma~\cite{Diggavi_Cover,Ihara}.
\begin{Lem}
\label{worst_additive} Let $\bbn$ be a Gaussian random vector with
covariance matrix $\bbsigma$, and $\bbk_X$ be a positive
semi-definite matrix. Consider the following optimization problem,
\begin{align}
\min_{p(\bx)}&\quad I(\bbn;\bbn+\bbx) \qquad {\rm s.t.}~~  {\rm
Cov}(\bbx)=\bbk_X
\end{align}
where $\bbx$ and $\bbn$ are independent. A Gaussian $\bbx$ is the
minimizer of this optimization problem.
\end{Lem}

We first bound $R_2$. Assume we fixed the distribution of
$(U,\bbx)$ such that ${\rm Cov}(\bbx)=\bbk_X$. Then, we have
\begin{align}
R_2 &\leq I(U;\bby^{2})-I(U;\bbz)\\
&=h(\bby^2)-h(\bbz)+[h(\bbz|U)-h(\bby^2|U)]\\
&\leq \frac{1}{2} \log
\frac{|\bbs+\bbsigma^2|}{|\bbs+\bbsigma_Z|}+[h(\bbz|U)-h(\bby^2|U)]
\label{worst_additive_noise_implies}
\end{align}
To show (\ref{worst_additive_noise_implies}), consider
$\tilde{\bbn}$ which is a Gaussian random vector with covariance
matrix $\bbsigma_Z-\bbsigma^2$, and is independent of
$(U,\bbx,\bbn^2)$. Thus, we can write
\begin{align}
h(\bby^2)-h(\bbz)&=h(\bbz|\tilde{\bbn})-h(\bbz)\\
&=-I(\tilde{\bbn};\bbx+\bbn^2+\tilde{\bbn})\\
&\leq \frac{1}{2} \log
\frac{|\bbk_X+\bbsigma^2|}{|\bbk_X+\bbsigma_Z|}\label{worst_additive_noise_implies1}\\
&\leq \frac{1}{2} \log \frac{|\bbs+\bbsigma^2|}{|\bbs+\bbsigma_Z|}
\label{monotonicity}
\end{align}
where (\ref{worst_additive_noise_implies1}) is due to
Lemma~\ref{worst_additive}, and (\ref{monotonicity}) follows from
the fact that
\begin{align}
\frac{|\bba|}{|\bba+\bbb|} \leq
\frac{|\bba+\bbdelta|}{|\bba+\bbb+\bbdelta|}
\end{align}
for $\bba\succeq \bzero, \bbb \succ \bzero, \bbdelta \succeq
\bzero$~\cite{MIMO_BC_Secrecy,Shamai_MIMO}.

For the rest of the proof, we need
Theorem~\ref{theorem_mimo_optimization}. According to
Theorem~\ref{theorem_mimo_optimization}, for any $(U,\bbx)$, there
exists a $\bzero\preceq \bbk\preceq {\rm Cov}(\bbx|U)$ such that
\begin{align}
h(\bbz|U)-h(\bby^2|U)&=\frac{1}{2} \log
\frac{|\bbk+\bbsigma_Z|}{|\bbk+\bbsigma^2|}\label{mimo_equality_1}\\
h(\bbz|U)-h(\bby^1_j|U)&\geq \frac{1}{2} \log
\frac{|\bbk+\bbsigma_Z|}{|\bbk+\bbsigma^1_j|},\quad j=1,\ldots,K_1
\label{mimo_inequality_1}
\end{align}
because $\bbsigma_j^1 \preceq \bbsigma^2,~j=1,\ldots,K_1$. Using
(\ref{mimo_equality_1}) in (\ref{worst_additive_noise_implies})
yields
\begin{align}
R_2 &\leq \frac{1}{2} \log
\frac{|\bbs+\bbsigma^2|}{|\bbk+\bbsigma^2|}-\frac{|\bbs+\bbsigma_Z|}{|\bbk+\bbsigma_Z|}
\end{align}
which is the desired bound on $R_2$.

The desired bound on $R_1$ can be obtained as follows
\begin{align}
R_1 & \leq \min_{j=1,\ldots,K_1}
I(\bbx;\bby^1_j|U)-I(\bbx;\bbz|U)\\
&=\min_{j=1,\ldots,K_1} h(\bby^1_j|U)-h(\bbz|U)-\frac{1}{2}\log
\frac{|\bbsigma^1_j|}{|\bbsigma_Z|} \\
&\leq  \min_{j=1,\ldots,K_1} \frac{1}{2} \log
\frac{|\bbk+\bbsigma_j^1|}{|\bbk+\bbsigma_Z|} -\frac{1}{2}\log
\frac{|\bbsigma^1_j|}{|\bbsigma_Z|}\label{mimo_inequality_implies}\\
&= \min_{j=1,\ldots,K_1} \frac{1}{2} \log
\frac{|\bbk+\bbsigma_j^1|}{|\bbsigma_j^1|}-\frac{1}{2} \log
\frac{|\bbk+\bbsigma_Z|}{|\bbsigma_Z|}
\end{align}
where (\ref{mimo_inequality_implies}) is due to
(\ref{mimo_inequality_1}). This completes the proof of
Theorem~\ref{theorem_mimo}.

\section{Proofs of Lemma~\ref{lemma_gaussian_optimization} and Theorem~\ref{theorem_partial}}
\label{proofs_for_partial_results}

\subsection{Proof of Lemma~\ref{lemma_gaussian_optimization}}
The optimization problem in (\ref{gaussian_optimization}) can be
put into the following alternative form
\begin{align}
\max_{\bzero \preceq \bbk \preceq \bbs}\quad  a+\mu b & \\
\textrm{s.t.}\quad
R_{1j}^{G}(\bbk) &\geq a,\quad j=1,\ldots,K_1 \\
R_{2k}^{G}(\bbk) &\geq b,\quad k=1,\ldots,K_2
\end{align}
which has the Lagrangian
\begin{align}
\mathcal{L}(\bbk)&=a+\mu b +\sum_{j=1}^{K_1} \lambda_{1j}
\left(R_{1j}^{G}(\bbk)-a \right)+\mu \sum_{k=1}^{K_2} \lambda_{2k}
\left(R_{2k}^{G}(\bbk)-b \right)+{\rm
tr}(\bbk\bbm)\nonumber\\
&\quad +{\rm tr}((\bbs-\bbk)\bbm_S)
\end{align}
where $\bbm$ and $\bbm_S$ are positive semi-definite matrices, and
$\{\lambda_{1j}\}_{j=1}^{K_1}$ and $\{\lambda_{2k}\}_{k=1}^{K_2}$
are non-negative. The KKT conditions are given by
\begin{align}
\frac{\partial \mathcal{L}(\bbk)}{\partial a}\Big|_{a=R_1^*}&=0 \label{KKT_1} \\
\frac{\partial \mathcal{L}(\bbk)}{\partial b}\Big|_{b=R_2^*}&=0 \label{KKT_2} \\
\nabla_{\bbk}\mathcal{L}(\bbk)|_{\bbk=\bbk^*}&=\bzero \label{KKT_3}\\
\lambda_{1j}(R_{1j}^{G}(\bbk^*)-R_1^{*})&=0,\qquad
j=1,\ldots,K_1 \label{KKT_4}\\
\lambda_{2k}(R_{2k}^{G}(\bbk^*)-R_2^{*})&=0,\qquad
k=1,\ldots,K_2\label{KKT_5}\\
{\rm tr}(\bbk^*\bbm)&=0 \label{KKT_6}\\
{\rm tr}\left((\bbs-\bbk^*)\bbm_S\right)&=0 \label{KKT_7}
\end{align}
The KKT conditions in (\ref{KKT_1}) and (\ref{KKT_2}) yield
$\sum_{j=1}^{K_1}\lambda_{1j}=1$ and
$\sum_{k=1}^{K_2}\lambda_{2k}=1$, respectively. Furthermore, the
KKT conditions in (\ref{KKT_4}) and (\ref{KKT_5}) imply
$\lambda_{1j}=0$ when $R_{1j}^G(\bbk^*)>R_1^*$ and
$\lambda_{2k}=0$ when $R_{2k}^G(\bbk^*)>R_2^*$, respectively. The
KKT condition in (\ref{KKT_3}) results in (\ref{KKT_lemma}).
Finally, since ${\rm tr}(\bba \bbb)={\rm tr}(\bbb \bba)\geq 0$
when $\bba\succeq \bzero$ and $\bbb\succeq \bzero$, we need to
have $\bbk^*\bbm=\bbm\bbk^*=\bzero$ and
$(\bbs-\bbk^*)\bbm_S=\bbm_S(\bbs-\bbk^*)=\bzero$.

\subsection{Proof of Theorem~\ref{theorem_partial}}

Let us fix $\{\lambda_{1j}\}_{j=1}^{K_1}$ and
$\{\lambda_{2k}\}_{k=1}^{K_2}$ as they are defined in
Lemma~\ref{lemma_gaussian_optimization}. We have
\begin{align}
\lefteqn{\max_{\bzero\preceq \bbk\preceq \bbs}
~\min_{j=1,\ldots,K_1}R_{1j}^{G}(\bbk)+\mu
\min_{k=1,\ldots,K_2}R_{2k}^{G}(\bbk) }\nonumber \\
&\leq \max_{(U,\bbx)} ~\min_{j=1,\ldots,K_1}R_{1j}+\mu
\min_{k=1,\ldots,K_2}R_{2k} \\
&\leq \max_{(U,\bbx)}~ \sum_{j=1}^{K_1}\lambda_{1j}
\left[I(\bbx;\bby_j^1|U)-I(\bbx;\bbz|U)\right]+\mu
\sum_{k=1}^{K_2}
\lambda_{2k} \left[I(U;\bby_k^2)-I(U;\bbz)\right] \\
&=\max_{(U,\bbx)}~ \sum_{j=1}^{K_1}\lambda_{1j}
\left[h(\bby_j^1|U)-h(\bbz|U)-\frac{1}{2}\log\frac{|\bbsigma_j^1|}{|\bbsigma_Z|}\right]+\mu
\sum_{k=1}^{K_2} \lambda_{2k} \left[h(\bby_k^2)-h(\bbz)\right]
\nonumber\\
&\qquad \qquad -\mu \sum_{k=1}^{K_2} \lambda_{2k}
\left[h(\bby_k^2|U)-h(\bbz|U)\right] \\
&\leq \max_{(U,\bbx)}~ \sum_{j=1}^{K_1}\lambda_{1j}
\left[h(\bby_j^1|U)-h(\bbz|U)-\frac{1}{2}\log\frac{|\bbsigma_j^1|}{|\bbsigma_Z|}\right]+\mu
\sum_{k=1}^{K_2} \lambda_{2k} \frac{1}{2} \log
\frac{|\bbs+\bbsigma_k^2|}{|\bbs+\bbsigma_Z|}
\nonumber\\
&\qquad \qquad -\mu \sum_{k=1}^{K_2} \lambda_{2k}
\left[h(\bby_k^2|U)-h(\bbz|U)\right] \label{supp_worst_additive} \\
&= \max_{\bzero \preceq \bbk \preceq \bbs}~
\sum_{j=1}^{K_1}\lambda_{1j} \left[\frac{1}{2}
\log\frac{|\bbk+\bbsigma_j^1|}{|\bbsigma_j^1|}-\frac{1}{2} \log
\frac{|\bbk+\bbsigma_Z|}{|\bbsigma_Z|}\right]\nonumber\\
&\qquad \qquad\qquad +\mu \sum_{k=1}^{K_2} \lambda_{2k} \left[
\frac{1}{2} \log
\frac{|\bbs+\bbsigma_k^2|}{|\bbk+\bbsigma_k^2|}-\frac{1}{2} \log
\frac{|\bbs+\bbsigma_Z|}{|\bbk+\bbsigma_Z|} \right] \label{tie_liu_optimization} \\
&=\max_{\bzero\preceq \bbk\preceq \bbs}
~\min_{j=1,\ldots,K_1}R_{1j}^{G}(\bbk)+\mu
\min_{k=1,\ldots,K_2}R_{2k}^{G}(\bbk) \label{kkt_lemma_implies}
\end{align}
where (\ref{supp_worst_additive}) comes from the fact that
\begin{align}
h(\bby_k^2)-h(\bbz)\leq \frac{1}{2} \log
\frac{|\bbs+\bbsigma_k^2|}{|\bbs+\bbsigma_Z|},\qquad
k=1,\ldots,K_2
\end{align}
which is a consequence of the worst additive lemma in
Lemma~\ref{worst_additive}, (\ref{tie_liu_optimization}) results
from Lemma~\ref{lemma_tie_liu}, (\ref{kkt_lemma_implies}) is due
to Lemmas~\ref{lemma_gaussian_optimization} and
\ref{lemma_tie_liu}. Thus, we have shown that
\begin{align}
\max_{\bzero\preceq \bbk\preceq \bbs}
~\min_{j=1,\ldots,K_1}R_{1j}^{G}(\bbk)+\mu
\min_{k=1,\ldots,K_2}R_{2k}^{G}(\bbk) = \max_{(U,\bbx)}
~\min_{j=1,\ldots,K_1}R_{1j}+\mu \min_{k=1,\ldots,K_2}R_{2k}
\end{align}
for $\mu\leq 1$, which completes the proof of theorem.

\section{Proof of Theorem~\ref{theorem_discrete_layered}}

\label{proof_of_theorem_discrete_layered}

We first show the achievability of the region given in
Theorem~\ref{theorem_discrete_layered}, then provide the converse
proof.

\subsection{Achievability}

We fix the distribution $p(u,x)$.

\vspace{0.5cm} \underline{\textbf{Codebook generation:}}
\begin{itemize}
\item Generate $2^{n(R_2+\tilde{R}_2)}$ length-$n$ $\bu$ sequences
through $p(\bu)=\prod_{i=1}^n p(u_i)$. Consider the permutation
$\pi_U$ on $\{1,\ldots,K_Z\}$ such that
\begin{align}
I(U;Z_{\pi_{U}(1)}) \leq \ldots \leq I(U;Z_{\pi_{U}(K_Z)})
\end{align}

We set $\tilde{R}_2$ as
\begin{align}
\tilde{R}_2 =\max_{t=1,\ldots,K_Z} I(U;Z_t)=I(U;Z_{\pi_U(K_Z)})
\label{dummy_message_rate_1}
\end{align}

We index $\bu$ sequences as
$\bu(w_2,\tilde{w}_{21},\ldots,\tilde{w}_{2K_Z})$ where
$w_2\in\{1,\ldots,2^{nR_2}\}$, and
$\tilde{w}_{2t}\in\{1,\ldots,2^{n\tilde{R}_{2t}}\},~t=1,\ldots,K_Z$.
$\tilde{R}_{2t}$ is given by
\begin{align}
\tilde{R}_{2t}=I(U;Z_{\pi_{U}(t)})-I(U;Z_{\pi_{U}(t-1)}),\quad
t=1,\ldots,K_Z \label{def_dummy_message_rate_1}
\end{align}
where we set $I(U;Z_{\pi_{U}(0)})=0$. We note that
\begin{align}
\sum_{t=1}^m \tilde{R}_{2t}=I(U;Z_{\pi_U (m)})
\end{align}
and in particular, for $m=K_Z$,
\begin{align}
\sum_{t=1}^{K_Z} \tilde{R}_{2t}=I(U;Z_{\pi_U
(K_Z)})=\max_{t=1,\ldots,K_Z} I(U;Z_t)=\tilde{R}_2
\end{align}

\item For each $\bu$, generate $2^{n(R_1+\tilde{R}_1)}$ length-$n$
$\bx$ sequences through $p(\bx|\bu)=\prod_{i=1}^n p(x_i|u_i)$.
Consider the permutation $\pi_X$ on $\{1,\ldots,K_2\}$ such that
\begin{align}
I(X;Y_{\pi_X(1)}^2|U)\leq \ldots \leq I(X;Y_{\pi_X(K_2)}^2|U)
\end{align}

We set $\tilde{R}_1$ as
\begin{align}
\tilde{R}_1=I(X;Y_{\pi_X(K_2)}^2|U)=\max_{k=1,\ldots,K_2}
I(X;Y_k^2|U) \label{dummy_message_rate_2}
\end{align}

We index $\bx$ sequences as
$\bx(w_1,\tilde{w}_{11},\ldots,\tilde{w}_{1K_2}|\bw_2)$ where
$\bw_2=(w_2,\tilde{w}_{21},\ldots,\tilde{w}_{2K_Z})$,
$w_1\in\{1,\ldots,2^{nR_1}\},$ and
$\tilde{w}_{1k}\in\{1,\ldots,2^{n\tilde{R}_{1k}}\},~k=1,\ldots,K_2$.
$\tilde{R}_{1k}$ is given by
\begin{align}
\tilde{R}_{1k}=I(X;Y^2_{\pi_X(k)}|U)-I(X;Y^2_{\pi_X(k-1)}|U),\quad
k=1,\ldots,K_2 \label{def_dummy_message_rate_2}
\end{align}
where we set $I(X;Y^2_{\pi_X(0)}|U)=0$. We note that
\begin{align}
\sum_{k=1}^m \tilde{R}_{1k}=I(X;Y^2_{\pi_X(m)}|U)
\end{align}
and in particular, for $m=K_2$, we have
\begin{align}
\sum_{k=1}^{K_2}
\tilde{R}_{1k}=I(X;Y_{\pi_X(K_2)}^2|U)=\max_{k=1,\ldots,K_2}I(X;Y^2_{k}|U)=\tilde{R}_1
\end{align}
\end{itemize}

\vspace{0.5cm} \underline{\textbf{Encoding:}}

\vspace{0.2cm} If $(w_1,w_2)$ is the message to be transmitted, we
pick $\{\tilde{w}_{1k}\}_{k=1}^{K_2}$ and
$\{\tilde{w}_{2t}\}_{t=1}^{K_Z}$ independently and uniformly, and
send the corresponding $\bx$.

\vspace{0.5cm} \underline{\textbf{Decoding:}}

\vspace{0.2cm}

The legitimate users can decode the messages with vanishingly
small probability of error, if the rates satisfy
\begin{align}
R_1+\tilde{R}_1& \leq \min_{j=1,\ldots,K_1} I(X;Y_{j}^1|U) \\
R_2+\tilde{R}_2& \leq \min_{k=1,\ldots,K_2} I(U;Y_{k}^2)
\end{align}
where we used the degradedness of the channel. Plugging the
expressions for $\tilde{R}_1$ and $\tilde{R}_2$ given in
(\ref{dummy_message_rate_1}) and (\ref{dummy_message_rate_2}), we
can get
\begin{align}
R_1& \leq \min_{\substack{j=1,\ldots,K_1\\ k=1,\ldots,K_2}} I(X;Y_{j}^1|U)- I(X;Y_k^2|U)\\
R_2& \leq \min_{\substack{k=1,\ldots,K_2\\t=1,\ldots,K_Z}}
I(U;Y_{k}^2)-I(U;Z_t)
\end{align}
which is the same as the region given in
Theorem~\ref{theorem_discrete_layered} because of the degradedness
of the channel.

\vspace{0.5cm} \underline{\textbf{Equivocation computation:}}

\vspace{0.2cm} We now show that this coding scheme satisfies the
secrecy requirements given in (\ref{perfect_secrecy_layered_1})
and (\ref{perfect_secrecy_layered_2}). We start with
(\ref{perfect_secrecy_layered_1})
\begin{align}
H(W_2|Z_{\pi_U(t)}^n)&=H(W_2,Z_{\pi_U(t)}^n)-H(Z_{\pi_U(t)}^n)\\
&=H(W_2,Z_{\pi_U(t)}^n,U^n)-H(U^n|W_2,Z_{\pi_U(t)}^n)-H(Z_{\pi_U(t)}^n)\\
&=H(U^n)+H(W_2,Z_{\pi_U(t)}^n|U^n)-H(U^n|W_2,Z_{\pi_U(t)}^n)-H(Z_{\pi_U(t)}^n)\\
&\geq H(U^n)-I(U^n;Z_{\pi_U(t)}^n)-H(U^n|W_2,Z_{\pi_U(t)}^n)
\label{equi_comp_1}
\end{align}
where we treat each term separately. Since $U^n$ can take
$2^{n(R_2+\tilde{R}_2)}$ values uniformly, for the first term, we
have
\begin{align}
H(U^n)=n(R_2+\tilde{R}_2) \label{equi_comp_2}
\end{align}
Following Lemma~8 of~\cite{Wyner}, the second term in
(\ref{equi_comp_1}) can be bounded as
\begin{align}
I(U^n;Z_{\pi_U(t)}^n)\leq nI(U;Z_{\pi_U(t)})+n\epsilon_{2,n}
\label{equi_comp_3}
\end{align}
where $\epsilon_{2,n}\rightarrow \infty$ as $n\rightarrow \infty$.
We now consider the third term of (\ref{equi_comp_1})
\begin{align}
H(U^n|W_2,Z_{\pi_U(t)}^n)&\leq
H(U^n,\tilde{W}_{2(t+1)},\ldots,\tilde{W}_{2K_Z}|W_2,Z_{\pi_U(t)}^n)\\
&\leq H(\tilde{W}_{2(t+1)},\ldots,\tilde{W}_{2K_Z})
+H(U^n|W_2,\tilde{W}_{2(t+1)},\ldots,\tilde{W}_{2K_Z},Z_{\pi_U(t)}^n)
\label{equi_comp_4_1}
\end{align}
The first term in (\ref{equi_comp_4_1}) is
\begin{align}
H(\tilde{W}_{2(t+1)},\ldots,\tilde{W}_{2K_Z})&=\sum_{l=t+1}^{K_Z}
H(\tilde{W}_{2l})\label{independence_etc_1}\\
&=\sum_{l=t+1}^{K_Z} n\tilde{R}_{2l} \label{uniform_etc}\\
&=n I(U;Z_{\pi_{U}(K_Z)})-nI(U;Z_{\pi_{U}(t)})
\label{equi_comp_4_2}
\end{align}
where (\ref{independence_etc_1}) is due to the independence of
$\{\tilde{W}_{2t}\}_{t=1}^{K_Z}$, (\ref{uniform_etc}) is due to
the fact that $\tilde{W}_{2t}$ can take $2^{n\tilde{R}_{2t}}$
values uniformly and independently for $t=1,\ldots,K_Z$, and in
(\ref{equi_comp_4_2}), we used the definitions of
$\{\tilde{R}_{2t}\}_{t=1}^{K_Z}$ given in
(\ref{def_dummy_message_rate_1}). We next consider the second term
in (\ref{equi_comp_4_1}). For that purpose, we note that given
\begin{align}
\left(~W_2=w_2,~\tilde{W}_{2(t+1)}=\tilde{w}_{2(t+1)},~\ldots~,~\tilde{W}_{2K_Z}=\tilde{w}_{2K_Z}~\right)
\label{side_info}
\end{align}
$U^n$ can take $2^{nI(U;Z_{\pi_U(t)})}$ values. Thus, given the
side information in (\ref{side_info}), the $\pi_U(t)$th
eavesdropper can decode $U^n$ with vanishingly small probability
of error, which implies that
\begin{align}
H(U^n|W_2,\tilde{W}_{2(t+1)},\ldots,\tilde{W}_{2K_Z},Z_{\pi_U(t)}^n)
\leq n\gamma_{2,n} \label{equi_comp_4_3}
\end{align}
due to Fano's lemma where $\gamma_{2,n}\rightarrow 0$ as
$n\rightarrow \infty$. Hence, plugging (\ref{equi_comp_4_2}) and
(\ref{equi_comp_4_3}) in (\ref{equi_comp_4_1}) yields
\begin{align}
H(U^n|W_2,Z_{\pi_U(t)}^n) \leq n I(U;Z_{\pi_{U}(K_Z)})-n
I(U;Z_{\pi_{U}(t)})+n\gamma_{2,n} \label{equi_comp_4}
\end{align}
Finally, using (\ref{equi_comp_2}), (\ref{equi_comp_3}) and
(\ref{equi_comp_4}) in (\ref{equi_comp_1}) yields
\begin{align}
H(W_2|Z_{\pi_U(t)}^n) &\geq n(R_2+\tilde{R}_2)-n\epsilon_{2,n}-n
I(U;Z_{\pi_{U}(K_Z)})-n\gamma_{2,n}\\
&=nR_2-n(\epsilon_{2,n}+\gamma_{2,n}) \label{equi_comp_1_1}
\end{align}
where we used (\ref{dummy_message_rate_1}). Since
(\ref{equi_comp_1_1}) implies (\ref{perfect_secrecy_layered_1}),
the proposed coding scheme ensures perfect secrecy for the second
group of users.

We now consider the second secrecy requirement given in
(\ref{perfect_secrecy_layered_2}).
\begin{align}
H(W_1|W_2,Y_{\pi_X(k)}^{2,n})&\geq
H(W_1|W_2,Y_{\pi_X(k)}^{2,n},U^n)\\
&=H(W_1|Y_{\pi_X(k)}^{2,n},U^n) \label{independence_etc_2}\\
&=H(W_1,Y_{\pi_X(k)}^{2,n}|U^n)-H(Y_{\pi_X(k)}^{2,n}|U^n)\\
&=H(X^n,W_1,Y_{\pi_X(k)}^{2,n}|U^n)-H(X^n|W_1,Y_{\pi_X(k)}^{2,n},U^n)-H(Y_{\pi_X(k)}^{2,n}|U^n)\\
&=H(X^n|U^n)+H(W_1,Y_{\pi_X(k)}^{2,n}|U^n,X^n)-H(X^n|W_1,Y_{\pi_X(k)}^{2,n},U^n)\nonumber\\
&\quad -H(Y_{\pi_X(k)}^{2,n}|U^n)\\
&\geq
H(X^n|U^n)-I(X^n;Y_{\pi_X(k)}^{2,n}|U^n)-H(X^n|W_1,Y_{\pi_X(k)}^{2,n},U^n)
\label{equi_comp_5}
\end{align}
where (\ref{independence_etc_2}) is due to the Markov chain
$W_2\rightarrow U^n \rightarrow (W_1,Y_{\pi_X(k)}^{2,n})$ which
originates from the coding scheme we proposed. Since given
$U^n=u^n$, $X^n$ can take $2^{n(R_1+\tilde{R}_1)}$ values
uniformly and independently, the first term in (\ref{equi_comp_5})
is
\begin{align}
H(X^n|U^n)=n(R_1+\tilde{R}_1) \label{equi_comp_6}
\end{align}
Following Lemma~8 of~\cite{Wyner}, the second term in
(\ref{equi_comp_5}) can be bounded as
\begin{align}
I(X^n;Y_{\pi_X(k)}^{2,n}|U^n) \leq
nI(X;Y_{\pi_X(k)}^{2}|U)+n\epsilon_{1,n} \label{equi_comp_7}
\end{align}
where $\epsilon_{1,n}\rightarrow 0$ as $n\rightarrow \infty$. We
now consider the third term in (\ref{equi_comp_5})
\begin{align}
H(X^{n}|W_1,U^n,Y_{\pi_X(k)}^{2,n})&\leq
H(X^{n},\tilde{W}_{1(k+1)},\ldots,\tilde{W}_{1K_2}|W_1,U^n,Y_{\pi_X(k)}^{2,n})\\
&\leq H(\tilde{W}_{1(k+1)},\ldots,\tilde{W}_{1K_2})
+H(X^{n}|W_1,U^n,Y_{\pi_X(k)}^{2,n},\tilde{W}_{1(k+1)},\ldots,\tilde{W}_{1K_2})
\label{equi_comp_8_1}
\end{align}
where the first term is given by
\begin{align}
H(\tilde{W}_{1(k+1)},\ldots,\tilde{W}_{1K_2})&=\sum_{l=k+1}^{K_2}
H(\tilde{W}_{1l})\label{independence_etc_3}\\
&=\sum_{l=k+1}^{K_2} n\tilde{R}_{1l} \label{uniform_etc_1}
\\
&=nI(X;Y_{\pi_X(K_2)}^2|U)-nI(X;Y_{\pi_X(k)}^2|U)
\label{equi_comp_8_2}
\end{align}
where (\ref{independence_etc_3}) is due to the independence of
$\{\tilde{W}_{1k}\}_{k=1}^{K_2}$, (\ref{uniform_etc_1}) comes from
the fact that $\tilde{W}_{1k}$ can take $2^{n\tilde{R}_{1k}}$
values uniformly and independently, and in (\ref{equi_comp_8_2}),
we used (\ref{def_dummy_message_rate_2}). We now bound the second
term of (\ref{equi_comp_8_1}). For that purpose, we first note
that given
\begin{align}
\left(~U^n=u^n,~W_1=w_1,~\tilde{W}_{1(k+1)}=\tilde{w}_{1(k+1)},~\ldots~,~\tilde{W}_{1K_2}=\tilde{w}_{1K_2}~\right)
\label{side_info_1}
\end{align}
$X^n$ can take $2^{nI(X;Y_{\pi_X(k)}^2|U)}$ values. Thus, given
the side information in (\ref{side_info_1}), the $\pi_X(k)$th user
in the second group can decode $X^n$ with vanishingly small
probability of error leading to
\begin{align}
H(X^{n}|W_1,U^n,Y_{\pi_X(k)}^{2,n},\tilde{W}_{1(k+1)},\ldots,\tilde{W}_{1K_2})\leq
n\gamma_{1,n}\label{equi_comp_8_3}
\end{align}
due to Fano's lemma where $\gamma_{1,n}\rightarrow 0$ as
$n\rightarrow \infty$. Plugging (\ref{equi_comp_8_2}) and
(\ref{equi_comp_8_3}) into (\ref{equi_comp_8_1}) yields
\begin{align}
H(X^{n}|W_1,U^n,Y_{\pi_X(k)}^{2,n})&\leq
nI(X;Y_{\pi_X(K_2)}^2|U)-nI(X;Y_{\pi_X(k)}^2|U)+n\gamma_{1,n}
\label{equi_comp_8}
\end{align}
Finally, using (\ref{equi_comp_6}), (\ref{equi_comp_7}) and
(\ref{equi_comp_8}) in (\ref{equi_comp_5}) results in
\begin{align}
H(W_1|W_2,Y_{\pi_X(k)}^{2,n})&\geq
nR_1+n\tilde{R}_1-nI(X;Y_{\pi_X(K_2)}^2|U)-n(\epsilon_{1,n}+\gamma_{1,n})\\
&=nR_1-n(\epsilon_{1,n}+\gamma_{1,n})
\end{align}
where we used (\ref{dummy_message_rate_2}). Since this implies
(\ref{perfect_secrecy_layered_2}), the proposed coding scheme
ensures perfect secrecy for the first group of users, completing
the proof.

\subsection{Converse}

First, we note that for an arbitrary code achieving the secrecy
rate pairs $(R_1,R_2)$, there exist
$(\epsilon_{1,n},\epsilon_{2,n})$ and
$(\gamma_{1,n},\gamma_{2,n})$ which vanish as $n\rightarrow
\infty$ such that
\begin{align}
H(W_1|Y_{j}^{1,n})&\leq n\epsilon_{1,n},\quad j=1,\ldots,K_1 \label{Fano_lemma_implies_l_1}\\
H(W_2|Y_{k}^{2,n})&\leq n\epsilon_{2,n},\quad k=1,\ldots,K_2 \label{Fano_lemma_implies_l_2}\\
I(W_2;Z_t^n)&\leq n\gamma_{2,n},\quad t=1,\ldots,K_Z
\label{perfect_secrecy_layered_1_implies}\\
I(W_1;Y_k^{2,n}|W_2)&\leq n\gamma_{1,n},\quad k=1,\ldots,K_2
\label{perfect_secrecy_layered_2_implies}
\end{align}
where (\ref{Fano_lemma_implies_l_1}) and
(\ref{Fano_lemma_implies_l_2}) are due to Fano's lemma, and
(\ref{perfect_secrecy_layered_1_implies}) and
(\ref{perfect_secrecy_layered_2_implies}) come from perfect
secrecy requirements in (\ref{perfect_secrecy_layered_1}) and
(\ref{perfect_secrecy_layered_2}).

We now define the following auxiliary random variables
\begin{align}
U_i=W_2 Y^{*,i-1} Z_{i+1}^{*,n},\quad i=1,\ldots,n
\end{align}
which satisfy the Markov chains
\begin{align}
U_i\rightarrow X_i \rightarrow Y_{j,i}^{1} \rightarrow Y_i^*
\rightarrow Y_{k,i}^2 \rightarrow Z_i^* \rightarrow Z_{t,i},\quad
i=1,\ldots,n \label{memoryless_channel}
\end{align}
for any $(j,k,t)$ triple. The Markov chain in
(\ref{memoryless_channel}) is a consequence of the fact that the
channel is memoryless and degraded.

We first establish the desired bound on $R_2$ as follows
\begin{align}
nR_2&=H(W_2)\\
&\leq I(W_2;Y_{k}^{2,n})+n\epsilon_{2,n}\\
&\leq I(W_2;Y_{k}^{2,n})-I(W_2;Z_{t}^{n})+n(\epsilon_{2,n}+\gamma_{2,n})\\
&= I(W_2;Y_{k}^{2,n}|Z_{t}^{n})+n(\epsilon_{2,n}+\gamma_{2,n})
\label{channel_is_degraded_l_1}\\
&= \sum_{i=1}^n I(W_2;Y_{k,i}^{2}|Z_{t}^{n},Y_{k}^{2,i-1})+n(\epsilon_{2,n}+\gamma_{2,n})\\
&= \sum_{i=1}^n
I(W_2;Y_{k,i}^{2}|Z_{t,i+1}^{n},Y_{k}^{2,i-1},Z_{t,i})+n(\epsilon_{2,n}+\gamma_{2,n})
\label{channel_is_degraded_l_2}\\
&\leq \sum_{i=1}^n
I(Z_{t,i+1}^{n},Y_{k}^{2,i-1},W_2;Y_{k,i}^{2}|Z_{t,i})+n(\epsilon_{2,n}+\gamma_{2,n})\\
&\leq \sum_{i=1}^n
I(Z_{i+1}^{*,n},Y^{*,i-1},Z_{t,i+1}^{n},Y_{k}^{2,i-1},W_2;Y_{k,i}^{2}|Z_{t,i})+n(\epsilon_{2,n}+\gamma_{2,n})\\
&\leq \sum_{i=1}^n
I(Z_{i+1}^{*,n},Y^{*,i-1},W_2;Y_{k,i}^{2}|Z_{t,i})+n(\epsilon_{2,n}+\gamma_{2,n})\label{channel_is_degraded_l_3}\\
&=\sum_{i=1}^n
I(U_i;Y_{k,i}^{2}|Z_{t,i})+n(\epsilon_{2,n}+\gamma_{2,n})
\end{align}
where (\ref{channel_is_degraded_l_1}) is due to the Markov chain
\begin{align}
W_2\rightarrow Y_{k}^{2,n}\rightarrow Z_{t}^{n}
\end{align}
which comes from the fact that the channel is degraded,
(\ref{channel_is_degraded_l_2}) results from the Markov chain
\begin{align}
Z_{t}^{i-1} \rightarrow Y_{k}^{2,i-1} \rightarrow
(W_2,Y_{k,i}^2,Z_{t,i}^n)
\end{align}
which is a consequence of the fact that the channel is memoryless
and degraded, and (\ref{channel_is_degraded_l_3}) is due to the
Markov chain
\begin{align}
(Z_{t,i+1}^{n},Y_{k}^{2,i-1})\rightarrow (
Z_{i+1}^{*,n},Y^{*,i-1})\rightarrow ( W_2,Y_{k,i}^{2},Z_{t,i})
\end{align}
which is a consequence of the Markov chain in
(\ref{our_Markov_chain}).

We now establish the bound on $R_1$ as follows
\begin{align}
nR_1&=H(W_1)\\
&=H(W_1|W_2) \\
&\leq I(W_1;Y_j^{1,n}|W_2)+n\epsilon_{1,n}\\
&\leq I(W_1;Y_j^{1,n}|W_2)-I(W_1;Y_k^{2,n}|W_2)+n(\epsilon_{1,n}+\gamma_{1,n})\\
&=I(W_1;Y_j^{1,n}|W_2,Y_k^{2,n})+n(\epsilon_{1,n}+\gamma_{1,n})\label{channel_is_degraded_l_4}\\
&=\sum_{i=1}^n
I(W_1;Y_{j,i}^{1}|W_2,Y_k^{2,n},Y_{j}^{1,i-1})+n(\epsilon_{1,n}+\gamma_{1,n})\\
&=\sum_{i=1}^n
I(W_1;Y_{j,i}^{1}|W_2,Y_{k,i+1}^{2,n},Y_{j}^{1,i-1},Y_{k,i}^2)+n(\epsilon_{1,n}+\gamma_{1,n})\label{channel_is_degraded_l_5}\\
&=\sum_{i=1}^n
I(W_1;Y_{j,i}^{1}|W_2,Y_{k,i+1}^{2,n},Y_{j}^{1,i-1},Z_{i+1}^{*,n},Y^{*,i-1},Y_{k,i}^2)+n(\epsilon_{1,n}+\gamma_{1,n})
\label{channel_is_degraded_l_6}\\
&=\sum_{i=1}^n
I(W_1;Y_{j,i}^{1}|U_i,Y_{k,i+1}^{2,n},Y_{j}^{1,i-1},Y_{k,i}^2)+n(\epsilon_{1,n}+\gamma_{1,n})\\
&\leq \sum_{i=1}^n
I(X_i,W_1;Y_{j,i}^{1}|U_i,Y_{k,i+1}^{2,n},Y_{j}^{1,i-1},Y_{k,i}^2)+n(\epsilon_{1,n}+\gamma_{1,n})\\
& =\sum_{i=1}^n
I(X_i;Y_{j,i}^{1}|U_i,Y_{k,i+1}^{2,n},Y_{j}^{1,i-1},Y_{k,i}^2)+n(\epsilon_{1,n}+\gamma_{1,n})
\label{channel_memoryless_4}\\
&=\sum_{i=1}^n
H(Y_{j,i}^{1}|U_i,Y_{k,i+1}^{2,n},Y_{j}^{1,i-1},Y_{k,i}^2)-H(Y_{j,i}^{1}|U_i,Y_{k,i+1}^{2,n},Y_{j}^{1,i-1},Y_{k,i}^2,X_i)\nonumber\\
&\quad +n(\epsilon_{1,n}+\gamma_{1,n})\\
&=\sum_{i=1}^n
H(Y_{j,i}^{1}|U_i,Y_{k,i+1}^{2,n},Y_{j}^{1,i-1},Y_{k,i}^2)-H(Y_{j,i}^{1}|U_i,Y_{k,i}^2,X_i)
+n(\epsilon_{1,n}+\gamma_{1,n}) \label{channel_memoryless_5}\\
&\leq \sum_{i=1}^n
H(Y_{j,i}^{1}|U_i,Y_{k,i}^2)-H(Y_{j,i}^{1}|U_i,Y_{k,i}^2,X_i)
+n(\epsilon_{1,n}+\gamma_{1,n})\label{conditioning_cannot_3}\\
&= \sum_{i=1}^n I(X_i;Y_{j,i}^{1}|U_i,Y_{k,i}^2)
+n(\epsilon_{1,n}+\gamma_{1,n})
\end{align}
where (\ref{channel_is_degraded_l_4}) is due to the Markov chain
\begin{align}
(W_1,W_2)\rightarrow Y_{j}^{1,n} \rightarrow Y_{k}^{2,n}
\end{align}
which comes from the degradedness of the channel,
(\ref{channel_is_degraded_l_5}) results from the Markov chain
\begin{align}
Y_{k}^{2,i-1}\rightarrow Y_{j}^{1,i-1} \rightarrow
(W_1,W_2,Y_{j,i}^{1},Y_{k,i}^{2,n})
\end{align}
which is again due to the degradedness of the channel,
(\ref{channel_is_degraded_l_6}) is a consequence of the Markov
chain
\begin{align}
(Z_{i+1}^{*,n},Y^{*,i-1})\rightarrow
(Y_{k,i+1}^{2,n},Y_{j}^{1,i-1})\rightarrow
(W_2,W_1,Y_{j,i}^1,Y_{k,i}^{2})
\end{align}
which results from the Markov chain in (\ref{our_Markov_chain}),
(\ref{channel_memoryless_4}) comes from the Markov chain
\begin{align}
(Y_{k,i}^{2},Y_{j,i}^1)\rightarrow X_i\rightarrow
(W_1,W_2,U_i,Y_{k,i+1}^{2,n},Y_{j}^{1,i-1})
\label{memoryless_channel_implies}
\end{align}
which is due to the fact that the channel is memoryless,
(\ref{channel_memoryless_5}) is also due to the Markov chain in
(\ref{memoryless_channel_implies}), and
(\ref{conditioning_cannot_3}) comes from the fact that
conditioning cannot increase entropy.

Single-letterization can be accomplished as outlined in the proofs
of Theorems~\ref{theorem_discrete} and
\ref{theorem_discrete_parallel}, completing the converse proof.

\section{Proof of Theorem~\ref{theorem_discrete_layered_parallel}}

\label{proof_of_theorem_discrete_layered_parallel}

The achievability of the region given in
Theorem~\ref{theorem_discrete_layered_parallel} can be shown by
selecting $(U,X)=(U_1,X_1,\ldots,U_L,X_L)$ with a joint
distribution of the form $p(u,x)=\prod_{\ell=1}^L
p(u_{\ell},x_{\ell})$. We next provide the converse proof. To that
end, we define the following auxiliary random variables
\begin{align}
U_{\ell,i}=W_2Y^{*,i-1}Z_{i+1}^{*,n}Y^*_{[1:\ell-1],i}Z^*_{[\ell+1:L],i},&\quad
i=1,\ldots,n, \quad \ell=1,\ldots,L
\end{align}
which satisfy the Markov chains
\begin{align}
U_{\ell,i}\rightarrow X_{\ell,i}\rightarrow Y_{j\ell,i}^1
\rightarrow Y^*_{\ell,i}\rightarrow Y_{k\ell,i}^2\rightarrow
Z_{\ell,i}^* \rightarrow Z_{t\ell,i},&\quad i=1,\ldots,n, \quad
\ell=1,\ldots,L 
\end{align}
for any $(j,k,t)$ triple. These Markov chains are a consequence of
the facts that the channel is memoryless and degraded, and
sub-channels are independent.

We first establish the desired bound on $R_2$. For that purpose,
following the proof of Theorem~\ref{theorem_discrete_layered}, we
get
\begin{align}
\lefteqn{nR_2\leq \sum_{i=1}^n
I(W_2;Y_{k,i}^2|Y_{k}^{2,i-1},Z_{t,i+1}^n,Z_{t,i})+n(\epsilon_{2,n}+\gamma_{2,n})}\\
&=\sum_{i=1}^n \sum_{\ell=1}^L
I(W_2;Y_{k\ell,i}^2|Y_{k}^{1,i-1},Z_{t,i+1}^n,Z_{t,i},Y_{k[1:\ell-1],i}^2)+n(\epsilon_{2,n}+\gamma_{2,n})\\
&=\sum_{i=1}^n \sum_{\ell=1}^L
I(W_2;Y_{k\ell,i}^2|Y_{k}^{2,i-1},Z_{t,i+1}^n,Z_{t[\ell+1:L],i},Y_{k[1:\ell-1],i}^2,Z_{t\ell,i})+n(\epsilon_{2,n}+\gamma_{2,n})
\label{degraded_1}\\
&\leq \sum_{i=1}^n \sum_{\ell=1}^L
I(Y^{*,i-1},Z_{i+1}^{*,n},Z_{[\ell+1:L],i}^*,Y_{[1:\ell-1],i}^*,Y_{k}^{2,i-1},Z_{t,i+1}^n,Z_{t[\ell+1:L],i},Y_{k[1:\ell-1],i}^2,W_2;Y_{k\ell,i}^2|Z_{t\ell,i})\nonumber\\
&\quad +n(\epsilon_{2,n}+\gamma_{2,n})\\
&=\sum_{i=1}^n \sum_{\ell=1}^L
I(Y^{*,i-1},Z_{i+1}^{*,n},Z_{[\ell+1:L],i}^*,Y_{[1:\ell-1],i}^*,W_2;Y_{k\ell,i}^2|Z_{t\ell,i})
+n(\epsilon_{2,n}+\gamma_{2,n})
\label{degraded_2}\\
&=\sum_{i=1}^n \sum_{\ell=1}^L
I(U_{\ell,i};Y_{k\ell,i}^2|Z_{t\ell,i})
+n(\epsilon_{2,n}+\gamma_{2,n})
\end{align}
where (\ref{degraded_1}) comes from the Markov chain
\begin{align}
Z_{t[1:\ell-1],i}\rightarrow Y_{k[1:\ell-1],i}^2\rightarrow
(W_2,Y_{k\ell,i}^2,Y_{k}^{2,i-1},Z_{t,i+1}^n,Z_{t[\ell:L],i})
\end{align}
which is a consequence of the facts that the channel is memoryless
and sub-channels are independent, (\ref{degraded_2}) results from
the Markov chain
\begin{align}
(Y_{k}^{2,i-1},Z_{t,i+1}^n,Z_{t[\ell+1:L],i},Y_{k[1:\ell-1],i}^2)\rightarrow
(Y^{*,i-1},Z_{i+1}^{*,n},Z_{[\ell+1:L],i}^*,Y_{[1:\ell-1],i}^*)\rightarrow
(W_2,Y_{k\ell,i}^2,Z_{t\ell,i})
\end{align}
which is a consequence of the Markov chain in
(\ref{our_Markov_chain_1}).

We now bound $R_1$. Following the proof of
Theorem~\ref{theorem_discrete_layered}, we get
\begin{align}
\lefteqn{nR_1 \leq \sum_{i=1}^n
I(W_1;Y_{j,i}^1|W_2,Y_{j}^{1,i-1},Y_{k,i+1}^{2,n},Y_{k,i}^2)+n
(\epsilon_{1,n}+\gamma_{1,n})} \\
&=\sum_{i=1}^n \sum_{\ell=1}^L
I(W_1;Y_{j\ell,i}^1|W_2,Y_{j}^{1,i-1},Y_{k,i+1}^{2,n},Y_{k,i}^2,Y_{j[1:\ell-1],i}^1)+n
(\epsilon_{1,n}+\gamma_{1,n})\\
&=\sum_{i=1}^n \sum_{\ell=1}^L
I(W_1;Y_{j\ell,i}^1|W_2,Y_{j}^{1,i-1},Y_{k,i+1}^{2,n},Y_{k[\ell+1:L],i}^2,Y_{j[1:\ell-1],i}^1,Y_{k\ell,i}^2)+n
(\epsilon_{1,n}+\gamma_{1,n})\label{degraded_3}\\
&=\sum_{i=1}^n \sum_{\ell=1}^L
I(W_1;Y_{j\ell,i}^1|U_{\ell,i},Y_{j}^{1,i-1},Y_{k,i+1}^{2,n},Y_{k[\ell+1:L],i}^2,Y_{j[1:\ell-1],i}^1,Y_{k\ell,i}^2)+n
(\epsilon_{1,n}+\gamma_{1,n})\label{degraded_4}\\
&\leq \sum_{i=1}^n \sum_{\ell=1}^L
I(X_{\ell,i},W_1;Y_{j\ell,i}^1|U_{\ell,i},Y_{j}^{1,i-1},Y_{k,i+1}^{2,n},Y_{k[\ell+1:L],i}^2,Y_{j[1:\ell-1],i}^1,Y_{k\ell,i}^2)+n
(\epsilon_{1,n}+\gamma_{1,n})\\
&= \sum_{i=1}^n \sum_{\ell=1}^L
I(X_{\ell,i};Y_{j\ell,i}^1|U_{\ell,i},Y_{j}^{1,i-1},Y_{k,i+1}^{2,n},Y_{k[\ell+1:L],i}^2,Y_{j[1:\ell-1],i}^1,Y_{k\ell,i}^2)+n
(\epsilon_{1,n}+\gamma_{1,n})\label{memoryless_channel_1} \\
&= \sum_{i=1}^n \sum_{\ell=1}^L
H(Y_{j\ell,i}^1|U_{\ell,i},Y_{j}^{1,i-1},Y_{k,i+1}^{2,n},Y_{k[\ell+1:L],i}^2,Y_{j[1:\ell-1],i}^1,Y_{k\ell,i}^2)\nonumber\\
&\quad
-H(Y_{j\ell,i}^1|U_{\ell,i},Y_{j}^{1,i-1},Y_{k,i+1}^{2,n},Y_{k[\ell+1:L],i}^2,Y_{j[1:\ell-1],i}^1,Y_{k\ell,i}^2,X_{\ell,i})+n
(\epsilon_{1,n}+\gamma_{1,n})\\
&= \sum_{i=1}^n \sum_{\ell=1}^L
H(Y_{j\ell,i}^1|U_{\ell,i},Y_{j}^{1,i-1},Y_{k,i+1}^{2,n},Y_{k[\ell+1:L],i}^2,Y_{j[1:\ell-1],i}^1,Y_{k\ell,i}^2)
-H(Y_{j\ell,i}^1|U_{\ell,i},Y_{k\ell,i}^2,X_{\ell,i}) \nonumber\\
&\quad +n(\epsilon_{1,n}+\gamma_{1,n}) \label{memoryless_channel_2} \\
&\leq \sum_{i=1}^n \sum_{\ell=1}^L
H(Y_{j\ell,i}^1|U_{\ell,i},Y_{k\ell,i}^2)
-H(Y_{j\ell,i}^1|U_{\ell,i},Y_{k\ell,i}^2,X_{\ell,i})+n(\epsilon_{1,n}+\gamma_{1,n})\label{conditioning_cannot_4}
\\
&=\sum_{i=1}^n \sum_{\ell=1}^L
I(X_{\ell,i};Y_{j\ell,i}^1|U_{\ell,i},Y_{k\ell,i}^2)+n(\epsilon_{1,n}+\gamma_{1,n})
\end{align}
where (\ref{degraded_3}) is due to the Markov chain
\begin{align}
Y_{k[1:\ell-1],i}^2\rightarrow Y_{j[1:\ell-1],i}^1 \rightarrow
(W_1,W_2,Y_{j}^{1,i-1},Y_{k,i+1}^{2,n},Y_{k[\ell:L],i}^2,Y_{j\ell,i}^1)
\end{align}
which is a consequence of the degradedness of the channel, and the
fact that sub-channels are independent and memoryless,
(\ref{degraded_4}) results from the Markov chain
\begin{align}
(Y^{*,i-1},Z_{i+1}^{*,n},Z_{[\ell+1:L],i}^*,Y_{[1:\ell-1],i}^*)\rightarrow
(Y_{j}^{1,i-1},Y_{k,i+1}^{2,n},Y_{k[\ell+1:L],i}^2,Y_{j[1:\ell-1],i}^1)
\rightarrow (W_1,W_2,Y_{j\ell,i}^1,Y_{k\ell,i}^2)
\end{align}
which is a consequence of the Markov chain in
(\ref{our_Markov_chain_1}), (\ref{memoryless_channel_1}) and
(\ref{memoryless_channel_2}) come from the Markov chain
\begin{align}
(W_1,U_{\ell,i},Y_{j}^{1,i-1},Y_{k,i+1}^{2,n},Y_{k[\ell+1:L],i}^2,Y_{j[1:\ell-1],i}^1)\rightarrow
X_{\ell,i} \rightarrow (Y_{k\ell,i}^2,Y_{j\ell,i}^1)
\end{align}
which is a consequence of the fact that sub-channels are
independent and memoryless.

We can obtain the desired single-letter expressions as it is done
in the proof of Theorem~\ref{theorem_discrete_parallel},
completing the proof.

\section{Proof of Theorem~\ref{theorem_parallel_layered_Gauss_optimization}}

\label{proof_of_theorem_parallel_layered_Gauss_optimization}

According to Theorem~\ref{theorem_parallel_Gauss_optimization},
there exists a $P^*\leq P$ such that
\begin{align}
h(X+\tilde{N}|U)-h(X+N^*|U)=\frac{1}{2} \log
\frac{P^*+\tilde{\sigma}^2}{P^*+\sigma_*^2}\label{dummy_opt_1}\\
h(X+\tilde{N}|U)-h(X+N_2|U)\leq \frac{1}{2} \log
\frac{P^*+\tilde{\sigma}^2}{P^*+\sigma_2^2}\label{dummy_opt_2}\\
h(X+\tilde{N}|U)-h(X+N_1|U)\geq \frac{1}{2} \log
\frac{P^*+\tilde{\sigma}^2}{P^*+\sigma_1^2} \label{dummy_opt_3}
\end{align}
for any $(\sigma_1^2,\sigma_2^2)$ as long as they satisfy
\begin{align}
\sigma_1^2 \leq \sigma_*^2 \leq \sigma_2^2 \leq \tilde{\sigma}^2
\end{align}

We first show (\ref{parallel_layered_opt_2}). To this end, we note
that (\ref{dummy_opt_1}) and (\ref{dummy_opt_2}) imply
\begin{align}
h(X+N_2|U)-h(X+N^*|U) \geq \frac{1}{2} \log
\frac{P^*+\sigma_2^2}{P^*+\sigma_*^2}
\label{parallel_layered_opt_1_1}
\end{align}
Furthermore, (\ref{dummy_opt_1}) and (\ref{dummy_opt_3}) imply
\begin{align}
h(X+N^*|U)-h(X+N_1|U)\geq \frac{1}{2} \log
\frac{P^*+\sigma_*^2}{P^*+\sigma_1^2}
\label{parallel_layered_opt_1_2}
\end{align}
Combining (\ref{parallel_layered_opt_1_1}) and
(\ref{parallel_layered_opt_1_2}) yields
\begin{align}
h(X+N_2|U)-h(X+N_1|U)\geq \frac{1}{2} \log
\frac{P^*+\sigma_2^2}{P^*+\sigma_1^2}
\end{align}
which is the desired result in (\ref{parallel_layered_opt_2}).

We now show (\ref{parallel_layered_opt_1}). We first note that we
can write $\tilde{N}$ as
\begin{align}
\tilde{N}=N_2+\sqrt{t}\tilde{N}_Z \label{stability_3}
\end{align}
where $\tilde{N}_Z$ is a zero-mean Gaussian random variable with
variance $\sigma_Z^2-\sigma_2^2$, and independent of $(U,X,N_2)$.
$t$ in (\ref{stability_3}) is given by
\begin{align}
t=\frac{\tilde{\sigma}^2-\sigma_2^2}{\sigma_Z^2-\sigma_2^2}
\label{def_t}
\end{align}
where it is clear that $t\in[0,1]$. We now use Costa's entropy
power inequality~\cite{Costa_EPI} to arrive at
(\ref{parallel_layered_opt_1})
\begin{align}
e^{2h(X+\tilde{N}|U)}&=e^{2h(X+N_2+\sqrt{t}\tilde{N}_Z|U)}\\
&\geq (1-t)e^{2h(X+N_2|U)}+te^{2h(X+N_Z|U)}
\end{align}
which is equivalent to
\begin{align}
e^{2\left[h(X+\tilde{N}|U)-h(X+N_2|U)\right]} &\geq
(1-t)+te^{2\left[h(X+N_Z|U)-h(X+N_2|U)\right]}
\end{align}
which can be written as
\begin{align}
h(X+N_Z|U)-h(X+N_2|U)&\leq \frac{1}{2} \log
\left[\frac{1}{t}e^{2\left[h(X+\tilde{N}|U)-h(X+N_2|U)\right]}-\frac{1-t}{t}\right]\\
&\leq \frac{1}{2} \log
\left[\frac{1}{t}\frac{P^*+\tilde{\sigma}^2}{P^*+\sigma_2^2}-\frac{1-t}{t}\right]\label{dummy_opt_2_implies}
\\
&= \frac{1}{2} \log
\left[\frac{P^*}{P^*+\sigma_2^2}-\frac{1}{t}\frac{\tilde{\sigma}^2-(1-t)\sigma_2^2}{P^*+\sigma_2^2}\right]\\
&= \frac{1}{2} \log
\frac{P^*+\sigma_Z^2}{P^*+\sigma_2^2}\label{def_t_implies}
\end{align}
where (\ref{dummy_opt_2_implies}) is due to (\ref{dummy_opt_2})
and (\ref{def_t_implies}) comes from (\ref{def_t}). Since
(\ref{def_t_implies}) is the desired result in
(\ref{parallel_layered_opt_1}), this completes the proof.

\section{Proof of Theorem~\ref{theorem_parallel_Gaussian_layered}}

\label{proof_of_theorem_parallel_Gaussian_layered}

Achievability is clear. We provide the converse proof. We fix the
distribution $\prod_{\ell=1}^L p(u_{\ell},x_{\ell})$ such that
\begin{align}
E\left[X_{\ell}^2\right]=P_{\ell},\quad \ell=1,\ldots,L
\end{align}
and $\sum_{\ell=1}^L P_{\ell}=P$. We first establish the bound on
$R_2$ given in (\ref{parallel_layered_R2}). To this end, we start
with (\ref{theorem_discrete_layered_parallel_R2}). Using the
Markov chain $U_{\ell}\rightarrow Y_{k\ell}^2\rightarrow
Z_{t\ell}$, we have
\begin{align}
R_2&\leq
\min_{\substack{k=1,\ldots,K_2\\t=1,\ldots,K_Z}}\sum_{\ell=1}^L
I(U_{\ell};Y_{k\ell}^2)-I(U_{\ell};Z_{t\ell})\label{thm_xx_implies}\\
&= \min_{\substack{k=1,\ldots,K_2\\t=1,\ldots,K_Z}}\sum_{\ell=1}^L
h(Y_{k\ell}^2)-h(Z_{t\ell})+\left[h(Z_{t\ell}|U_{\ell})-h(Y_{k\ell}^2|U_{\ell})\right]\\
&\leq
\min_{\substack{k=1,\ldots,K_2\\t=1,\ldots,K_Z}}\sum_{\ell=1}^L
\frac{1}{2} \log
\frac{P_{\ell}+\Lambda_{k,\ell\ell}^2}{P_{\ell}+\Lambda_{t,\ell\ell}^Z}+\left[h(Z_{t\ell}|U_{\ell})-h(Y_{k\ell}^2|U_{\ell})\right]
\label{epi_implies_1}
\end{align}
where (\ref{epi_implies_1}) comes from the fact that
\begin{align}
h(Y_{k\ell}^2)-h(Z_{t\ell})
\end{align}
is maximized by Gaussian distribution which can be shown by using
the entropy power inequality~\cite{Stam,Blachman}. We now use
Theorem~\ref{theorem_parallel_layered_Gauss_optimization}. For
that purpose, we introduce $\bblambda_Y^*$ and $\bblambda_Z^*$
which satisfy
\begin{align}
\bblambda_j^1 \preceq \bblambda_Y^* \preceq \bblambda_k^2 \preceq
\bblambda_Z^* \preceq \bblambda_t^Z
\end{align}
for any $(j,k,t)$ triple, and in particular, for the diagonal,
elements of these matrices, we have
\begin{align}
\Lambda_{j,\ell\ell}^1 \leq \Lambda_{Y,\ell\ell}^* \leq
\Lambda_{k,\ell\ell}^2 \leq \Lambda_{Z,\ell\ell}^* \leq
\Lambda_{t,\ell\ell}^Z
\end{align}
for any $(j,k,t,\ell)$. Thus, due to
Theorem~\ref{theorem_parallel_layered_Gauss_optimization}, for any
selection of $\{(U_{\ell},X_{\ell})\}_{\ell=1}^L$, we have
\begin{align}
P_\ell^* &\leq P_\ell \label{thm_xx_implies_1} \\
h(Z_{t\ell}|U_{\ell})-h(Y_{k\ell}^2|U_{\ell})&\leq \frac{1}{2}
\log
\frac{P_{\ell}^*+\Lambda_{t,\ell\ell}^Z}{P_{\ell}^*+\Lambda_{k,\ell\ell}^2} \label{thm_xx_implies_2}\\
h(Y_{k\ell}^2|U_{\ell})-h(Y_{j\ell}^1|U_{\ell})&\geq \frac{1}{2}
\log
\frac{P_{\ell}^*+\Lambda_{k,\ell\ell}^2}{P_{\ell}^*+\Lambda_{j,\ell\ell}^1}
\label{thm_xx_implies_3}
\end{align}
for any $(k,j,t,\ell)$. Using (\ref{thm_xx_implies_2}) in
(\ref{epi_implies_1}) yields
\begin{align}
R_2 &\leq
\min_{\substack{k=1,\ldots,K_2\\t=1,\ldots,K_Z}}\sum_{\ell=1}^L
\frac{1}{2} \log
\frac{P_{\ell}+\Lambda_{k,\ell\ell}^2}{P_{\ell}^*+\Lambda_{k,\ell\ell}^2}-\frac{1}{2}
\log
\frac{P_{\ell}+\Lambda_{t,\ell\ell}^Z}{P_{\ell}^*+\Lambda_{t,\ell\ell}^Z}
\end{align}
By defining $P_{\ell}^*=\beta_{\ell} P_{\ell}$ and
$\bar{\beta}_{\ell}=1-\beta_{\ell},~\ell=1,\ldots,L$, where
$\beta_{\ell}\in[0,1]$ due to (\ref{thm_xx_implies_1}), we get the
desired bound on $R_2$ given in (\ref{parallel_layered_R2}).

We now bound $R_1$. We start with
(\ref{theorem_discrete_layered_parallel_R1}). Using the Markov
chain $U_{\ell}\rightarrow X_{\ell}\rightarrow
Y_{j\ell}^1\rightarrow Y_{k\ell}^2$, we have
\begin{align}
R_1&\leq \min_{\substack{j=1,\ldots,K_1\\k=1,\ldots,K_2}}
\sum_{\ell=1}^L
I(X_{\ell};Y_{j\ell}^1|U_{\ell})-I(X_{\ell};Y_{k\ell}^2|U_{\ell})
\label{thm_yy_implies} \\
&= \min_{\substack{j=1,\ldots,K_1\\k=1,\ldots,K_2}}
\sum_{\ell=1}^L
h(Y_{j\ell}^1|U_{\ell})-h(Y_{k\ell}^2|U_{\ell})-\frac{1}{2}\log\frac{\Lambda_{j,\ell\ell}^1}{\Lambda_{k,\ell\ell}^2}\\
&\leq \min_{\substack{j=1,\ldots,K_1\\k=1,\ldots,K_2}}
\sum_{\ell=1}^L \frac{1}{2}
\log\frac{P_{\ell}^*+\Lambda_{j,\ell\ell}^1}{P_{\ell}^*+\Lambda_{k,\ell\ell}^2}-\frac{1}{2}\log\frac{\Lambda_{j,\ell\ell}^1}{\Lambda_{k,\ell\ell}^2}
\label{thm_xx_implies_3_1}\\
&=\min_{\substack{j=1,\ldots,K_1\\k=1,\ldots,K_2}} \sum_{\ell=1}^L
\frac{1}{2}
\log\left(1+\frac{\beta_{\ell}P_{\ell}}{\Lambda_{j,\ell\ell}^1}\right)
-\frac{1}{2}\log\left(1+\frac{\beta_{\ell}P_{\ell}}{\Lambda_{k,\ell\ell}^2}\right)
\label{thm_xx_implies_3_2}
\end{align}
where (\ref{thm_xx_implies_3_1}) is due to
(\ref{thm_xx_implies_3}). Since (\ref{thm_xx_implies_3_2}) is the
desired bound on $R_1$ given in (\ref{parallel_layered_R1}), this
completes the proof.

\section{Background Information for Appendix~\ref{proof_of_theorem_mimo_layered_optimization}}

\label{background_information}

In Appendix~\ref{proof_of_theorem_mimo_layered_optimization}, we
need some properties of the Fisher information and the
differential entropy, which are provided here.

\begin{Def}[\!\!\cite{MIMO_BC_Secrecy}, Definition~3]
Let $(\bbu,\bbx)$ be an arbitrarily correlated length-$n$ random
vector pair with well-defined densities. The conditional Fisher
information matrix of $\bbx$ given $\bbu$ is defined as
\begin{align}
\bbj(\bbx|\bbu)=E\left[\brho(\bbx|\bbu)\brho(\bbx|\bbu)^\top\right]
\end{align}
where the expectation is over the joint density $f(\bu,\bx)$, and
the conditional score function $\brho(\bx|\bu)$ is
\begin{align}
\brho(\bx|\bu)=\nabla \log f(\bx|\bu)=\left[~\frac{\partial\log
f(\bx|\bu)}{\partial x_1}~~\ldots~~\frac{\partial\log
f(\bx|\bu)}{\partial x_n}~\right]^\top
\end{align}
\end{Def}

The following lemma will be used in the upcoming proof. In fact,
an unconditional version of this lemma is proved in Lemma 6
of~\cite{MIMO_BC_Secrecy}.

\begin{Lem}
\label{lemma_change_in_fisher} Let $\bbt,\bbu,\bbv_1,\bbv_2$ be
random vectors such that $(\bbt,\bbu)$ and $(\bbv_1,\bbv_2)$ are
independent. Moreover, let $\bbv_1,\bbv_2$ be Gaussian random
vectors with covariances matrices $\bbsigma_1,\bbsigma_2$ such
that $\bzero \prec \bbsigma_1 \preceq \bbsigma_2$. Then, we have
\begin{align}
\bbj^{-1}(\bbu+\bbv_2|\bbt)-\bbsigma_2 \succeq
\bbj^{-1}(\bbu+\bbv_1|\bbt)-\bbsigma_1
\end{align}
\end{Lem}

The following lemma is also instrumental for the upcoming proof
whose proof can be found in~\cite{MIMO_BC_Secrecy}.

\begin{Lem}[\!\!\cite{MIMO_BC_Secrecy},~Lemma~8]
\label{Shamai_s_lemma} Let $\bbk_1,\bbk_2$ be positive
semi-definite matrices satisfying
$\bzero\preceq\bbk_1\preceq\bbk_2$, and $\mathbf{f}(\bbk)$ be a
matrix-valued function such that $\mathbf{f}(\bbk)\succeq\bzero$
for $\bbk_1\preceq\bbk\preceq \bbk_2$. Then, we have
\begin{align}
\int_{\bbk_1}^{\bbk_2}\mathbf{f}(\bbk)d\bbk \geq 0
\end{align}
\end{Lem}

The following generalization of the de Bruin identity~\cite{Stam,
Blachman} is due to~\cite{Palomar_Gradient}.
In~\cite{Palomar_Gradient}, the unconditional form of this
identity, i.e., the case where $U=\phi$, is proved. However, its
generalization to this conditional form for an arbitrary $U$ is
rather straightforward, and given in Lemma~16
of~\cite{MIMO_BC_Secrecy}.

\begin{Lem}[\!\!\cite{MIMO_BC_Secrecy}, Lemma~16]
\label{gradient_fisher_conditional} Let $(\bbu,\bbx)$ be an
arbitrarily correlated random vector pair with finite second order
moments, and be independent of the random vector $\bbn$ which is
zero-mean Gaussian with covariance matrix $\bbsigma_N\succ\bzero$.
Then, we have
\begin{align}
\nabla_{\bbsigma_N} h(\bbx+\bbn|\bbu)=\frac{1}{2}
\bbj(\bbx+\bbn|\bbu)
\end{align}
\end{Lem}

\section{Proof of Theorem~\ref{theorem_mimo_layered_optimization}}

\label{proof_of_theorem_mimo_layered_optimization}

According to Theorem~\ref{theorem_mimo_optimization}, for any
selection of $(U,\bbx)$, there exists a $\bbk^*\preceq \bbs$ such
that
\begin{align}
h(\bbx+\bbn^*|U)-h(\bbx+\bbn_2|U)&=\frac{1}{2} \log
\frac{|\bbk^*+\bbsigma^*|}{|\bbk^*+\bbsigma_2|} \label{thm_zz_implies_1}\\
h(\bbx+\bbn^*|U)-h(\bbx+\bbn_1|U)&\geq \frac{1}{2} \log
\frac{|\bbk^*+\bbsigma^*|}{|\bbk^*+\bbsigma_1|}
\label{thm_zz_implies_2}
\end{align}
for any $\bbsigma_1$ such that $\bbsigma_1\preceq \bbsigma_2$.
Furthermore, $\bbk^*$ satisfies~\cite{MIMO_BC_Secrecy}
\begin{align}
\bbk^*\preceq \bbj^{-1}(\bbx+\bbn^*|U)-\bbsigma^*
\label{thm_zz_implies_3}
\end{align}
Equations (\ref{thm_zz_implies_1}) and (\ref{thm_zz_implies_2})
already imply
\begin{align}
h(\bbx+\bbn_2|U)-h(\bbx+\bbn_1|U)&\geq \frac{1}{2} \log
\frac{|\bbk^*+\bbsigma_2|}{|\bbk^*+\bbsigma_1|}
\end{align}
for any $\bbsigma_1$ such that $\bbsigma_1\preceq \bbsigma_2$,
which is the desired inequality in
(\ref{mimo_layered_inequality_2}).

We now prove (\ref{mimo_layered_inequality_1}). For that purpose,
we note that (\ref{thm_zz_implies_3}) implies
\begin{align}
\bbk^*\preceq \bbj^{-1}(\bbx+\bbn|U)-\bbsigma_N
\label{thm_zz_implies_3_1}
\end{align}
for any Gaussian random vector $\bbn$, independent of $(U,\bbx)$,
with covariance matrix $\bbsigma_N$ such that $\bbsigma_N \succeq
\bbsigma^*$ because of Lemma~\ref{lemma_change_in_fisher}. The
order in (\ref{thm_zz_implies_3_1}) is equivalent to
\begin{align}
\bbj(\bbx+\bbn|U)\preceq (\bbk^*+\bbsigma_N)^{-1},\quad \bbsigma^*
\preceq \bbsigma_N \label{thm_zz_implies_3_2}
\end{align}

Now, we can obtain (\ref{mimo_layered_inequality_1}) as follows
\begin{align}
h(\bbx+\bbn_Z|U)-h(\bbx+\bbn_2|U)&=h(\bbx+\bbn_Z|U)-h(\bbx+\bbn^*|U)\nonumber\\
&\quad +h(\bbx+\bbn^*|U)-h(\bbx+\bbn_2|U)\\
&=h(\bbx+\bbn_Z|U)-h(\bbx+\bbn^*|U)+\frac{1}{2}
\log\frac{|\bbk^*+\bbsigma^*|}{|\bbk^*+\bbsigma_2|}\label{smthng_implies_1}
\\
&=\frac{1}{2}\int_{\bbsigma^*}^{\bbsigma_Z}\bbj(\bbx+\bbn|U)~d\bbsigma_N+\frac{1}{2}
\log\frac{|\bbk^*+\bbsigma^*|}{|\bbk^*+\bbsigma_2|}\label{smthng_implies_2}
\\
&\leq
\frac{1}{2}\int_{\bbsigma^*}^{\bbsigma_Z}(\bbk^*+\bbsigma_N)^{-1}
d\bbsigma_N+\frac{1}{2}
\log\frac{|\bbk^*+\bbsigma^*|}{|\bbk^*+\bbsigma_2|}\label{smthng_implies_3}\\
&\leq \frac{1}{2}
\log\frac{|\bbk^*+\bbsigma_Z|}{|\bbk^*+\bbsigma_2|}\label{smthng_implies_4}
\end{align}
where (\ref{smthng_implies_1}) is due to (\ref{thm_zz_implies_1}),
(\ref{smthng_implies_2}) is obtained by using
Lemma~\ref{gradient_fisher_conditional}, and
(\ref{smthng_implies_3}) comes from Lemma~\ref{Shamai_s_lemma} by
noting (\ref{thm_zz_implies_3_2}). Since (\ref{smthng_implies_4})
is the desired inequality in (\ref{mimo_layered_inequality_1}),
this completes the proof.

\section{Proof of Theorem~\ref{theorem_mimo_layered}}

\label{proof_of_theorem_mimo_layered}

We first establish the desired bound on $R_2$ given in
(\ref{mimo_layered_R2}) as follows
\begin{align}
R_2&\leq \min_{t=1,\ldots,K_Z} I(U;\bby^2)-I(U;\bbz_t)\label{thm_gg_implies_1}\\
&=\min_{t=1,\ldots,K_Z}
h(\bby^2)-h(\bbz_t)+\left[h(\bbz_t|U)-h(\bby^2|U)\right]\\
&\leq \min_{t=1,\ldots,K_Z} \frac{1}{2}
\log\frac{|\bbs+\bbsigma^2|}{|\bbs+\bbsigma_t^Z|}+\left[h(\bbz_t|U)-h(\bby^2|U)\right]
\label{worst_additive_noise_implies2}
\end{align}
where (\ref{thm_gg_implies_1}) comes from
Theorem~\ref{theorem_discrete_layered} by noting the Markov chain
$U\rightarrow \bby^2\rightarrow \bbz_t$, and
(\ref{worst_additive_noise_implies2}) can be obtained by using the
worst additive noise lemma, i.e., Lemma~\ref{worst_additive}, as
it is done in the proof of Theorem~\ref{theorem_mimo}. We now use
Theorem~\ref{theorem_mimo_layered_optimization}. According to
Theorem~\ref{theorem_mimo_layered_optimization}, for any selection
of $(U,\bbx)$, there exists a positive semi-definite matrix $\bbk$
such that $\bbk\preceq \bbs$ and
\begin{align}
h(\bbz_t|U)-h(\bby^2|U) &\leq \frac{1}{2}
\log\frac{|\bbk+\bbsigma_t^Z|}{|\bbk+\bbsigma^2|} \label{thm_gg_implies_2} \\
h(\bby^2|U)-h(\bby_j^1|U) &\geq \frac{1}{2}
\log\frac{|\bbk+\bbsigma^2|}{|\bbk+\bbsigma_j^1|}
\label{thm_gg_implies_3}
\end{align}
for any $(j,t)$ pair. Using (\ref{thm_gg_implies_2}) in
(\ref{worst_additive_noise_implies2}) yields
\begin{align}
R_2&\leq \min_{t=1,\ldots,K_Z} \frac{1}{2}
\log\frac{|\bbs+\bbsigma^2|}{|\bbk+\bbsigma^2|}-\frac{1}{2}
\log\frac{|\bbs+\bbsigma_t^Z|}{|\bbk+\bbsigma_t^Z|}
\end{align}
which is the desired bound on $R_2$ given in
(\ref{mimo_layered_R2}).

We now obtain the desired bound on $R_1$ given in
(\ref{mimo_layered_R1}) as follows
\begin{align}
R_1&\leq
\min_{j=1,\ldots,K_1}I(\bbx;\bby^1_j|U)-I(\bbx;\bby^2|U)\label{thm_gg_implies_4}\\
&=\min_{j=1,\ldots,K_1} h(\bby^1_j|U)-h(\bby^2|U)-\frac{1}{2}
\log\frac{|\bbsigma_j^1|}{|\bbsigma^2|} \\
&\leq \min_{j=1,\ldots,K_1} \frac{1}{2}
\log\frac{|\bbk+\bbsigma_j^1|}{|\bbsigma_j^1|}-\frac{1}{2}
\log\frac{|\bbk+\bbsigma^2|}{|\bbsigma^2|}\label{thm_gg_implies_5}
\end{align}
where (\ref{thm_gg_implies_4}) comes from
Theorem~\ref{theorem_discrete_layered} by noting the Markov chain
$U\rightarrow \bbx\rightarrow \bby_j^1\rightarrow \bby^2$ and
(\ref{thm_gg_implies_5}) is obtained by using
(\ref{thm_gg_implies_3}). Since (\ref{thm_gg_implies_5}) is the
desired bound on $R_1$ given in (\ref{mimo_layered_R1}), this
completes the proof.

\bibliographystyle{unsrt}
\bibliography{IEEEabrv,references2}
\end{document}